\documentclass[10pt,conference]{IEEEtran}
\usepackage[left=0.67in,right=0.67in,top=0.75in,bottom=1.05in]{geometry}
\IEEEoverridecommandlockouts
\usepackage{cite, balance}
\usepackage{amsmath,amssymb,amsfonts}
\usepackage{enumitem}
\usepackage{graphicx}
\usepackage{textcomp}
\usepackage{subcaption}
\usepackage{dblfloatfix}
\usepackage[utf8]{inputenc}
\usepackage{url}
\usepackage{xcolor}
\usepackage{color}
\usepackage{multirow}
\usepackage{url}
\usepackage{algpseudocode}
\usepackage{algorithm}
\usepackage[T1]{fontenc}
\usepackage{newtxtext,newtxmath}
\usepackage{diagbox}
\usepackage{hhline}
\usepackage{multirow}
\def\BibTeX{{\rm B\kern-.05em{\sc i\kern-.025em b}\kern-.08em
    T\kern-.1667em\lower.7ex\hbox{E}\kern-.125emX}}

\newcommand{\pgm}[1]{{\textcolor{black}{#1}}}
\newcommand{\bigtablefont}{\fontsize{11}{12}\selectfont}

\begin{document}
%\title{Adaptive Energy Efficient LEO Satellite \\ Optical Feeder Link Scheduling for Delay-Tolerant Traffic \\}
\title{Energy Efficient Traffic Scheduling\\ For Optical LEO Satellite Downlinks\\}

\author{
\IEEEauthorblockN{
Ethan Fettes\IEEEauthorrefmark{1},
Pablo G. Madoery\IEEEauthorrefmark{1},
Halim Yanikomeroglu\IEEEauthorrefmark{1},
Gunes Karabulut Kurt\IEEEauthorrefmark{1}\IEEEauthorrefmark{6},}
Abhishek Naik\IEEEauthorrefmark{2},
Stéphane Martel\IEEEauthorrefmark{4}\\
%Khaled Ahmed\IEEEauthorrefmark{4}

\vspace*{0.3cm}

\IEEEauthorblockA{\IEEEauthorrefmark{1} Non-Terrestrial Networks (NTN) Lab, Department of Systems and Computer Engineering, Carleton University, Canada}
\IEEEauthorblockA{\IEEEauthorrefmark{6} Poly-Grames Research Center, Department of Electrical Engineering, Polytechnique Montréal, Montréal, Canada}
\IEEEauthorblockA{\IEEEauthorrefmark{2} National Research Council Canada, Canada}
\IEEEauthorblockA{\IEEEauthorrefmark{4} Satellite Systems, MDA, Canada}

}

\maketitle
\IEEEpubidadjcol

\begin{abstract}
In recent years, the number of satellites in orbit has increased rapidly, with megaconstellations like Starlink providing near-global, delay-sensitive communication services. However, not all satellite communication use cases have stringent delay requirements; services such as Earth observation (EO) and remote Internet of Things (IoT) fall into this category. These relaxed delay quality of service (QoS) objectives allow services to be delivered using sparse constellations, enabled by delay-tolerant networking protocols. In the context of rapidly growing data volumes that must be delivered through satellite networks, a key challenge is having sufficient \pgm{space-to-ground link capacity}. This has led to proposals for using free-space optical (FSO) communications, which offer high data rates. However, FSO communications are highly vulnerable to weather-related disruptions. This results in certain communication opportunities being energy inefficient. Given the energy-constrained nature of satellites, developing schemes to improve energy efficiency is highly desirable. In this work, both static and adaptive schemes were developed to balance maintaining the delivery ratio and maximizing energy efficiency. The proposed schemes fall into the following categories: threshold schemes, heuristic sorting algorithms, and reinforcement learning-based schemes. The schemes were evaluated under a variety of different data volumes and cloud cover distribution configurations \pgm{as well as a case study using historical weather data}. It was found that static schemes suffered from low delivery ratio performance under dynamic conditions when compared to adaptive techniques. However, this performance improvement came at the cost of increased complexity and onboard computations.
 
\end{abstract}


\section{Introduction}
As the space economy grows, a wide variety of quality of service (QoS) requirements from different applications must be serviced by satellite networks \pgm{\cite{bs_no1}}. Although much attention has been given to traffic with delay-sensitive QoS requirements associated with devices such as smartphones, delay-tolerant traffic is also growing with the advent of paradigms such as direct-IoT \cite{Sparse_Constellations} and increasing Earth observation (EO) data volumes \cite{EO_Data_Volume}. Unlike delay-sensitive traffic, delay-tolerant traffic does not have stringent delay QoS requirements, with hours or even days of delay being acceptable depending on the specific application. This provides the opportunity for sparse networks, which require fewer satellites and thus cost less than megaconstellations.

To facilitate the efficient transmission of data through sparse satellite networks, delay-tolerant networking (DTN) protocols have been developed, enabling nodes to store and forward data efficiently without requiring continuous end-to-end connectivity between the sender and receiver. 

Due to the aforementioned high demand for EO and satellite IoT (SIoT) services, the data volume that must be delivered through sparse constellations is rapidly increasing. A key bottleneck in these architectures is the link between the satellite and the ground. To address this issue, free-space optical (FSO) communication systems have been proposed. However, FSO communication systems are heavily impacted by adverse weather conditions, such as cloud cover. As cloud cover can be forecast using weather models, such as the Global Forecast System (GFS) \cite{Cloud_Cover_bias}, or machine learning-based techniques \cite{cloud_prediction}, this information can be used to make intelligent downlink decisions.

During communication windows between satellites and ground stations, establishing FSO links requires coarse pointing, acquisition, and other processes to enable successful communication \cite{ATP_Survey}. All of these services require energy provided by the spacecraft's power system. Unnecessary energy expenditures are undesirable due to the power-limited nature of spacecraft  \cite{Power_Budget_Operations}. Hence, developing strategies to use more favourable transmission periods (frequently referred to as contacts) is desirable for satellite operators.

%Limitations of existing works
\pgm{Although existing works have explored optical downlink scheduling in the context of fully-connected delay-sensitive satellite networks, such as in  \cite{Lyu1}, \cite{Lyu3}, there has been very little work in the context of delay-tolerant networks.}% The downlink scheduling problem in this case, is well-studied in the RF context \cite{RF_Downlink_Scheduling}, but is largely unexplored in the optical context.}

In this work, static and adaptive schemes that use cloud cover forecast data to improve energy efficiency while maximizing the delivered customer data were developed and tested. The static schemes follow set strategies that remain unchanged during the downlink period. The adaptive schemes can adjust the transmission logic during the downlink period. All proposed schemes were tested using different configurations to determine their performance in various key scenarios, including situations where changes to the ground infrastructure and data volume requirements occur. \pgm{ In addition, a case study was conducted using a realistic ground station deployment location, satellite orbit configuration, and historical weather data. This case study allowed for validation of the results obtained in the more abstract simulations and provided insights into future work and limitations of the proposed downlink scheduling schemes.} These efforts resulted in the following contributions:

\begin{itemize}
    \item We show that the optical downlink scheduling problem can be formulated as a knapsack problem variant.
    \item Develop static and adaptive energy-efficient optical downlink schemes for a delay-tolerant traffic in a sparse satellite network.
    \item \pgm{Demonstrate the benefits of using adaptive scheduling schemes, specifically designed for delay-tolerant satellite networks}.
    \item Discuss future work and considerations when implementing the proposed algorithms.
\end{itemize}

The rest of the paper is structured as follows. Section \ref{sec:background} provides an overview of relevant background and related work in this domain, followed by Section \ref{sec:systemmodel}, which outlines the system model and shows how the problem can be formulated as a knapsack problem variant. Section \ref{sec:proposed_schemes} provides descriptions of all proposed static and adaptive schemes. Section \ref{sec:Sim_settings} describes the specific training and testing parameters of the proposed schemes. Section \ref{sec:evaluation} presents the overall performance of the proposed schemes and evaluates the impact of changes in cloud cover and data volume distribution. Lastly, Sections \ref{conclusion} and \ref{futurework} discuss important takeaways from the previous sections and outline potential avenues for future research.

\section{Background and Related Work}
\label{sec:background}
%We provide background regarding the movement to LEO satellites
\pgm{To provide context to the challenges briefly mentioned in the previous section, an overview of DTN in satellite networks and the challenges of space-to-ground optical links will be provided. Subsequently, we discuss existing network planning strategies to mitigate weather-induced disruptions to the space-to-ground link and examine the limitations of these approaches. Finally, we introduce the current state-of-the-art optical downlink scheduling approaches.}

\subsection{Delay-Tolerant Networking and Key Applications}
DTN protocols have been developed for a wide variety of scenarios, including both terrestrial and satellite networks \cite{CGRTutorial}. Unlike in terrestrial networks, where communication opportunities between sparse nodes are stochastic, in satellite networks, these opportunities, known as contacts, are largely deterministic.

Historically, the most important use case of DTN for satellite networks has been deep-space applications; however, in recent years, DTN has been proposed for near-Earth applications such as satellite IoT (SIoT) and EO \cite{Sparse_Constellations}. 

The most popular DTN protocol for satellite networks is bundle protocol, which packages data into bundles \cite{CGRTutorial}. Store-and-forward capability is provided by the bundle layer on top of other protocols, such as TCP/IP. It enables nodes to store data until the next transmission opportunity.

To enable efficient routing in sparse networks, specialized routing protocols are required due to the lack of end-to-end connectivity and frequent topology changes. To address these challenges, contact graph routing (CGR) has been developed for delay-tolerant routing in satellite networks. CGR leverages the predictability of satellite orbits to represent sparse networks as a graph of contacts efficiently.

%CGR with uncertainty
A limitation of standard CGR is that it treats all contacts between network nodes as deterministic \cite{CGR_Uncert}, and it does not explicitly account for traffic or specific QoS requirements. This results in suboptimal performance in various scenarios. To address these drawbacks, variants of CGR have been proposed.

In \cite{CGR_Uncert}, an uncertainty framework is used to evaluate delay-centric, reliability-centric, and energy efficiency-centric CGR variants under uncertainty. To directly address the problem of uncertain contacts in CGR, \cite{MDP_CGR} proposes a Markov decision process (MDP) modelling solution to determine the optimal policy within this framework. However, this framework only considers on-off contact failure, whereas many contacts may experience intermittent failure within a single contact. In addition, as this approach assumes perfect knowledge of the environment's transition dynamics, it can only be considered as a theoretical upper bound on performance.
%Gaps and why using these techniques is not suitable
%Environment dynamics must be known
%Unnecessary complexity for simple satellite systems
%Does not take into account specific environmental factors related to optical downlink
%No specific energy saving policy just a side effect

Although these proposed CGR variants offer notable benefits in meeting specific QoS requirements and handling uncertain contacts, they increase complexity, and some variants do not provide benefits for all network topologies \cite{CGR_Uncert}. 

\subsection{Weather Impairments in Space-to-Ground FSO communications}

 Space-to-ground FSO links are affected by two main effects: Mie scattering and geometric scattering \cite{FSO_link_budget}. Mie scattering is due to cloud cover, while geometrical scattering is associated with fog, dust, and other low-altitude weather conditions. Frequently, attenuation due to cloud cover vastly exceeds the link margin, with values exceeding 100 dB/km possible \cite{Ground_Station_Optimization}. Hence, the presence of cloud cover in the line of sight between the satellite and the ground station can be assumed to cause link failure. The availability of a link in the presence of cloud cover can be represented as a fractional value between zero and one \cite{Lyu1}.

Although the link availability for a given cloud cover value is highly stochastic, cloud cover conditions themselves can be predicted to a certain degree of accuracy using numerical weather models. However, due to the limitations of current numerical weather models, cloud cover forecasts can have a significant amount of uncertainty associated with the output, with medium-range cloud cover forecasts from the global forecast system (GFS) having a mean absolute error of over 20 percent \cite{Cloud_Cover_bias}. This limits the temporal length of a downlink period, as noted in \cite{reactive_planning}, to several days at most. The fractional nature of link availability and the feasible duration of a downlink period significantly impact the nature of downlink scheduling strategies utilized in DTN applications.

\subsection{Optical Ground Station Network Planning and Architecture Solutions}
 
 For space-to-ground FSO links, the most common method for mitigating the challenge of adverse weather conditions is to make use of site diversity \cite{Site_Diversity_Clouds}. However, this approach fails to address the issue of using energy-inefficient contacts and is expensive to implement. 

In \cite{gsNetOptimization}, the optical ground station placement problem is analyzed using historical weather data. By considering the correlation between ground station availability, an efficient algorithm is designed to optimally place ground stations to maximize visibility of geostationary Earth orbit (GEO) satellites.

 If using multiple ground station sites is not possible to achieve site diversity, novel network architectures, such as high-altitude platform system (HAPS)-enhanced ground stations, are also a potential alternative \cite{ethanspaper}. However, these solutions come with high capital costs that, in many DTN circumstances, may not be justified or feasible. This highlights the need for downlink scheduling strategies that improve optical downlink performance without requiring new infrastructure.
\subsection{Downlink Scheduling and Weather-Aware Techniques}
Without investing in new ground-based infrastructure, optical downlink scheduling aims to optimize the use of existing satellite-to-ground contacts. In the following section, existing downlink scheduling approaches will be reviewed, and research gaps will be identified.
%RF downlink and difference with Optical downlink

\pgm {Satellite downlink scheduling is a well-studied problem in RF communications, with particular focus on EO applications. In \cite{RF_Downlink_Scheduling}, this problem is explored in the context of the RADARSAT-2 satellite mission. The authors identify the satellite scheduling problem as NP-hard and propose heuristic-based solutions. }

\pgm{In \cite{EO_RF_Optical}, an advanced heuristic algorithm based on tabu search is developed to solve the downlink scheduling problem for a combined RF-Optical system. In terms of the system model, satellites are assumed to have both optical and RF terminals to transmit data, for reliability purposes. Although this paper shows good performance in the delay-tolerant optical satellite downlink case, it does not consider the unreliable nature of the optical links due to adverse weather conditions.}

%German Reactive Optical Downlink paper
%\pgm{To properly consider adverse weather conditions in the satellite downlink scheduling problem,} accurate cloud cover forecasts. These forecasts can be provided by conventional short and medium-term weather forecast models such as the GFS \cite{reactive_planning}. This information can be fed to appropriate algorithms as part of a larger downlink scheduling system to meet the needs of a particular satellite system. %A key consideration in a generic downlink scheduling system is the allocation of optical ground stations when multiple satellites are competing for downlink opportunities. This results in the need to implement fairness, particularly when the number of satellites greatly exceeds the number of ground stations (otherwise, this effect does not have a significant impact).

In \cite{Lyu1}, optical downlink scheduling is considered in the context of a satellite system with multiple GEOs and optical ground stations. The objective is to maximize the data volume delivered in the current time period, given a set of cloud cover conditions. The authors proposed a greedy heuristic algorithm to solve this problem by sorting space-to-ground contacts according to a metric informed by stochastic link availability. This algorithm was evaluated in terms of its computational complexity when compared to optimal solutions.

This work is extended in \cite{Lyu3}, which considers multi-source uncertainty in downlink scheduling, specifically the uncertainty in both link availability and data transfer rates. To accomplish this, robust optimization is used to formulate the problem, and a robust downlink scheduling algorithm is proposed. Although deterministic algorithms can achieve higher throughput than the proposed robust algorithm, they do not offer the same performance guarantees.

Downlink scheduling provides notable performance benefits when applied in isolation, as noted in \cite{Lyu3} and \cite{Lyu1}. Applying it in coordination with site diversity strategies is also highly advantageous. In \cite{Lyu2}, downlink scheduling is applied in coordination with the optical ground station placement problem. The objective defined by the authors is to maximize the data throughput of a multi-layer optical satellite network while maintaining compliance with a total cost constraint for the ground station infrastructure. The scheduling and ground station placement problem is combined to jointly optimize four decision variables: low Earth orbit (LEO) downlink scheduling; GEO downlink scheduling; ground station deployment locations, and the number of laser terminals per ground station. This problem was solved using a branch-and-bound algorithm, and the authors demonstrated that it achieved superior performance compared to existing methods. This demonstrates that traditional approaches, such as site diversity, also benefit from considering optical downlink scheduling.\\

%Put the research gaps here that we seek to fill
%Although \cite{reactive_planning} provides a comprehensive overview of a generic system for planning optical downlinks, it does not consider what algorithms could be potentially used within this system in pursuit of performance objectives. 

Several algorithms are presented in \cite{Lyu1} and \cite{Lyu2}, including branch-and-bound and heuristic algorithms. These algorithms are specifically designed to maximize data throughput in a delay-sensitive network context. Considering the case of a delay-tolerant satellite system, unlike delay-sensitive approaches, algorithms can adjust in real-time during the downlink period due to the temporal nature of the problem. Downlink schemes for these delay-tolerant networks are developed in this work, with a focus on sparse LEO networks and energy efficiency.

%Furthermore, considering energy efficiency, given its importance in satellite operations, could lead to interesting solutions that are not considered in the purely data throughput case.  

\section{System Model and Problem Formulation}
\label{sec:systemmodel}
\pgm{In the following section, the system model is defined, including the weather and satellite system architecture. Subsequently, the problem objectives are introduced, and it is shown how the problem can be formulated as a knapsack problem variant.}

\subsection{System Architecture}
\label{sec:system_Arc}
\begin{figure}
\centering
\includegraphics[width=0.98\linewidth]{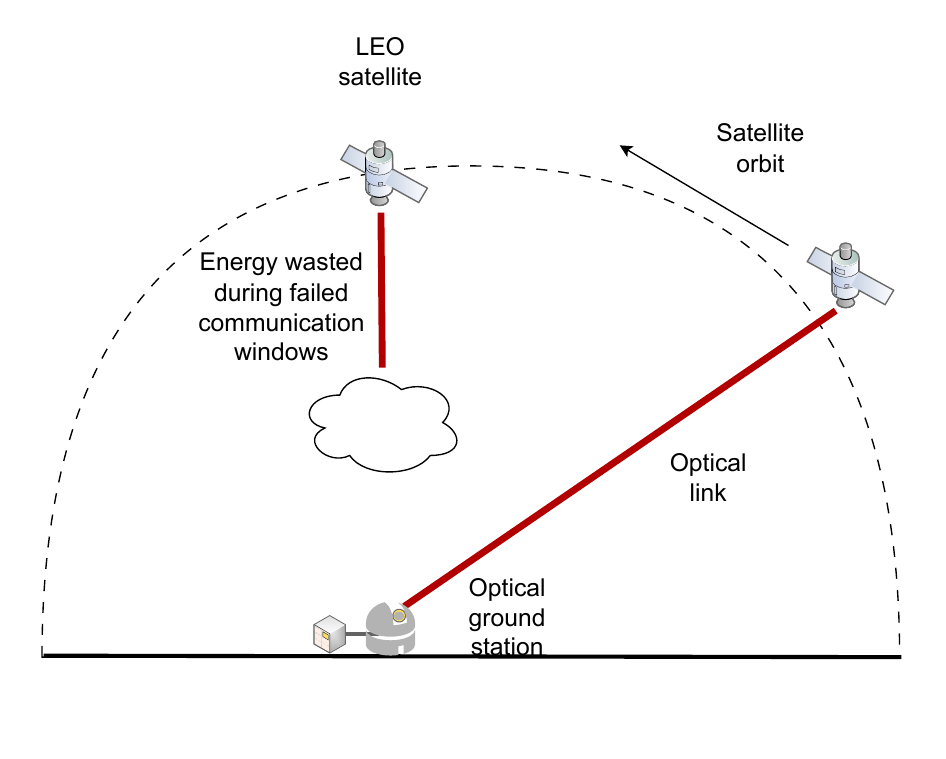} 
  \caption{System architecture.}
  \label{fig:Sys_Arch}
\end{figure}

%Figure with satellite and 2 ground stations
The system model shown in Figure \ref{fig:Sys_Arch} outlines the key aspects of the efficient optical downlink scheduling problem. For the purposes of this work, a single LEO satellite is used as the space segment of the network. \pgm{Due to the orbital dynamics associated with LEO satellites, the satellite will only have intermittent connectivity with the ground stations \cite{bs_no2} }. A downlink period is defined as a set time period during which it is expected that all delay-tolerant traffic present on the satellite will be delivered to the ground. It is assumed that all the delay-tolerant traffic to be delivered is present on the satellite before the downlink period. Additionally, it is also assumed that only one ground station is available at any one time. %Given this architecture, the data must be stored and forwarded by the satellite, necessitating the use of delay-tolerant networking (DTN) protocols.
\begin{table}[ht]
\centering
\caption{List of Symbols}
\label{tab:symbols}
\begin{tabular}{ll}
\hline
\textbf{Symbol} & \textbf{Description} \\
\hline
$e_i$ & Energy Expenditure of the $i^{th}$ Contact\\
$v_i$& Contact Volume of the $i^{th}$ Contact\\
$d_i$ & Packets Delivered during the $i^{th}$ Contact\\
$V$ & Total Contact Volume\\
$\mathsf{CC}_i$& Cloud Cover of Contact of the $i^{th}$ Contact  \\
$\mathsf{CA}_i$ & Contact Availability of the $i^{th}$ Contact\\
$\mathsf{lr}$& Link Sample Rate\\
$\mathsf{DV}$ & Packets Remaining to Deliver during Downlink Period\\
$\mathsf{dr}$ & Data Rate\\
$\mathsf{pl}$ & Packet Length\\
%$ls$ & Link Sample\\
$\mathsf{DR}$ & Contact Delivery Ratio\\
$\mathsf{DV}_{init}$ & Initial Number of Packets to Deliver\\
$x_i$ & Decision Variable of the $i^{th}$ Contact\\
$\mathsf{EE}$ & Energy Efficiency\\
$N$ & Number of Contacts in Downlink Period\\
$w$ & Problem Objective Weight\\
$N$ & Number of Contacts\\
$\mathsf{cv}_i$ & Contact Value of the $i^{th}$ Contact\\
$\mathsf{wt}_i$ & Contact Weight of $i^{th}$ Contact\\
$W$ & Total Contact Weight Constraint\\
$\beta$ & Objective Weight\\
$\upsilon$ & Objective Weight \\
$n$ & Current Contact Iterator Value\\
$T$ & Threshold Value\\
$\mathsf{nTe}$& Number of Test Episodes For Each Setting\\
$\mathsf{Tgr}$ & Threshold Step Granularity\\
$\mathsf{nTv}$ & Number of Threshold Values to Test\\
$\mathsf{Tsr}$ & Threshold Scheme Delivery Ratio\\
$\mathsf{Bsr}$ & Baseline Scheme Delivery Ratio\\
$\pgm{\mathsf{\tau}}$ & \pgm{Volume Margin}\\
$\mathsf{SGC}$ & Space-to-Ground Contacts \\
$\mathsf{CVR}$ & Contact Volume Remaining\\
$\mathsf{SC}$ & Sorted Contacts\\
$E$ & Total Energy Expenditure\\
$D$ & Expected Data Volume to Deliver\\
$X$ & Matrix Containing Contact Decisions\\
$\mathsf{SC}_2$ & Auxiliary Sorted Contact Matrix\\
$D_2$ & Auxiliary Data Volume\\
$c$ & Reward Scaling Factor \\
$\mathsf{DR}$ & Total Delivery Ratio\\
$\mathsf{CT}$ & Total Contact Time\\
$\mathsf{ER}$ & Terminal Reward\\
$\mathsf{SR}$ & Step Reward\\
$R$ & Total Return\\
$m$ & Episode Max Step Index\\
$\epsilon$ & Exploration Parameter\\
%Q & Action-Value\\
%a & Action\\
%S & State\\
%S' & Next State\\
%O & Observation\\
%O' & Next Observation\\
$\alpha$ & Learning Rate\\
$\gamma$ & Discount Factor\\
\hline
\end{tabular}
\end{table}
%The most popular DTN protocol for satellite networks is packet protocol, which packages data into packets \cite{CGRTutorial}. The use of DTN protocols is necessary, as traditional internet protocols assume a fully-connected network with small delays between nodes. If a network is not fully connected such as in the case of a sparse satellite network, these protocols will not function properly. Store and forward capability is provided by the packet layer on top of other protocols, such as IP and TCP and allows nodes to store data until the next transmission opportunity. To allow for efficient routing in these networks, contact graph routing (CGR) is a popular choice. CGR takes advantage of the predictable nature of satellite orbits to represent sparse networks efficiently in terms of a graph of contacts.\\

%Weather Modelling
To model the impact of adverse weather conditions, the space-to-ground communication channel is modelled as an on-off channel. The channel status, and hence the availability of a ground station during a contact, depend on the satellite's position relative to the clouds and the ground station. This position changes rapidly due to the satellite orbit. The link availability value for contact $i$ is denoted $\mathsf{CA}_i$. This value is calculated using the logic defined in Algorithm \ref{pseudocode:Link_Avail} where the inputs are the cloud cover $\mathsf{CC}_i$, the contact volume $v_i$ and the link sample rate $\mathsf{lr}$.

The $v_i$ is the total amount of data that can be sent during a satellite-to-ground contact. The longer a space-to-ground contact lasts, the higher its contact volume is. In terms of units, the contact volume in this paper is measured in packets.
\begin{algorithm}
\caption{Link Availability model}
\label{pseudocode:Link_Avail}
\begin{algorithmic}[1]
    \State $\mathsf{CC} \gets \text{Cloud Cover}$
    \State $v \gets \text{Contact Volume}$
    \State $\mathsf{CA} \gets 0$
    \State $\mathsf{lr} \gets \text{Link Sample Rate}$
    \For{$k \gets 1 \ \textbf{to} \ v  \mathsf{lr}$}
        \State $\mathsf{ls} \gets \text{Uniform}(0,1)$
        \If{$\mathsf{CC}<\mathsf{ls} $}
            \State $\mathsf{CA} \gets \mathsf{CA}+1$
        \EndIf
    \EndFor
    \State $Output \gets \mathsf{CA}$
\end{algorithmic}
\end{algorithm}
Given a certain $\mathsf{CA}$ value, the number of packets delivered during a contact is defined in Algorithm \ref{pseudocode:Dat_Transfer}, which takes the $\mathsf{CA}_i$, the remaining packets stored in the satellite buffer $\mathsf{DV}$, the packet length $\mathsf{pl}$, the link data rate $\mathsf{dr}$, the contact volume $v_i$ and the link sample rate $\mathsf{lr}$. The output of this algorithm is the number of delivered packets $d_i$, $\mathsf{DV}$ and the excess energy expenditure $e_i$.
\begin{algorithm}
    \caption{Data Transfer Model}
    \label{pseudocode:Dat_Transfer}
    \begin{algorithmic}[1]
        \State $d \gets 0$
        \State $e \gets 0$
        \State $\mathsf{DV} \gets \text{Remaining Data}$
        \State $\mathsf{CA} \gets \text{Contact Availability}$
        \State $v \gets \text{Contact Volume}$
        \State $\mathsf{lr} \gets \text{Link Sample Rate}$
        \State $\mathsf{dr} \gets \text{Data Rate}$
        \State $\mathsf{pl} \gets \text{Packet Size}$
        \For{$i \gets 1 \  \textbf{to} \ \mathsf{CA}$}
            \If{$\mathsf{DV}>0$}
                \If{$\mathsf{DV} >= \frac{\mathsf{dr}/\mathsf{lr}}{\mathsf{pl}}$}
                    \State $\mathsf{DV} \gets \mathsf{DV}-\frac{\mathsf{dr}/\mathsf{lr} }{\mathsf{pl}}$
                    \State $d \gets d+\frac{\mathsf{dr}/\mathsf{lr}}{\mathsf{pl}}$
                \Else
                    \State $d \gets d+\mathsf{DV}$
                    \State $\mathsf{DV} \gets 0$
                \EndIf
            \EndIf
        \EndFor
        \State $e \gets v-d$
        \State $Output \gets d,e,\mathsf{DV}$
        
    \end{algorithmic}
\end{algorithm}
As spacecraft power usage modelling is outside the scope of this work, a simple model is used to measure excess power consumption. The power model is limited to communications power rather than total spacecraft power. We define excess power consumption as the number of unsuccessful packet transmissions due to cloud cover. For a given contact $i$, the  excess power consumption $e_i$ can be defined as
\begin{equation}
e_i=v_i-d_i.
\end{equation}
This assumes that the acquisition and pointing processes required to re-acquire communication between the satellite and the ground station consume spacecraft power constantly during the link-failure periods. Excess power consumption can be eliminated in the event of a link failure by disabling the laser terminal pointing and acquisition functions; however, this also eliminates the ability to transmit if conditions improve during the contact. Finally, it is assumed that an acknowledgement mechanism is present, notifying transmitting nodes when packets are received.

In terms of how Algorithm \ref{pseudocode:Link_Avail} and Algorithm \ref{pseudocode:Dat_Transfer} fit within the problem dynamics, potential solutions do not have access to the $\mathsf{CA}_i$ for a given contact $i$; they only view the weather conditions, the expended energy and the number of delivered packets. If all the data allocated for the downlink period is transmitted, then any subsequent time periods will only result in increased excess energy expenditure.
\subsection{Problem Formulation}
\label{sec:problem_form}
\paragraph*{Objectives}
Given the system model described above, the two key metrics of interest are delivery ratio and energy expenditure. The overall delivery ratio for a downlink period can be expressed as
\begin{equation}
    \mathsf{DR}=\frac{\sum_{i=1}^N d_ix_i}{\mathsf{DV}_{init}},
\end{equation}
where $x$ is a decision variable that indicates if a contact is selected, and is defined as
\begin{equation}
    x \in \{0,1\}.
\end{equation}
As noted previously, after all packets are delivered, no further packets can be delivered with only further energy expenditure incurred (including if a contact is only partially utilized). Furthermore, the total excess energy expenditure $E$ incurred over a downlink period can be expressed as 
\begin{equation}
    E=\sum_{i=1}^N e_ix_i.
\end{equation}

%where $d$ is the number of delivered packets according to Algorithm \ref{pseudocode:Link_Avail}.

%Energy efficiency $y$ is defined as the number of packets delivered, divided by the contact volume in packets. \\

%\begin{equation}
%y=1-\frac{\sum_{i=1}^Na_i}{\sum_{i=1}^Nx_ib_i},\text{ if } %\sum_{i=1}^Nx_i>0
    %\begin{cases}
    %undefined, \text{ if } \sum_{i=1}^Nx_i<=0\\
    %\end{cases}
%\end{equation}

The algorithms defined in subsequent sections were designed under the assumption that the delivery ratio is the most important metric for satellite operators, as delivering the maximum amount of sensor data is highly desirable. The secondary objective is to improve energy efficiency. However, all algorithms can be easily modified to increase energy expenditure performance, at the expense of a decrease in delivery ratio performance. These objectives can be represented as follows:
%\begin{equation}
%    Z=\alpha \times w+(1-\alpha)\times y,
%\end{equation}
%Need to make an overall problem objective statement
%\begin{equation}
%    max \ w \frac{ \sum_{i=1}^N d_ix_i-max(0,\sum_{i=1}^N d_ix_i-DV_{init})}{\sum_{i=1}^Nv_ix_i}+(1-w) \sum_{i=1}^Nx_i -\sum_{i=1}^N \frac{v_i-d_i}{v_i}x_i +\frac{max(0,\sum_{i=1}^Nd_ix_i-DV)}{}
%\end{equation}

%\begin{equation}
%    max \ w\sum_{i=1}^N  CDR_i\times x_i+\frac{(1- w)}{\sum_{i=1}^Nx_i}\sum_{i=1}^NEE_i \times x_i,
%\end{equation}

%\begin{align}
%    \text{s.t.} \quad & \text{C1: } \sum_{i=1}^N d_i \leq DV_{init}, \\
%                      & \text{C2: } w \in [0, 1],\\
%                      & \text{C3: } \sum_{i=1}^Nx_i>0,
%\end{align}
\begin{equation}
    max \ w \sum_{i=1}^Nd_ix_i-(1-w)\sum_{i=1}^Ne_ix_i
\end{equation}
\begin{align}
    \text{s.t.} \quad & \text{C1: } \sum_{i=1}^N d_i \leq \mathsf{DV_{init}}, \\
                      & \pgm{\text{C2: } w \in [0, 1]}
                      %& \text{C3: } \sum_{i=1}^Nx_i>0,
\end{align}
where $w$ is a weight value that indicates the importance placed on the respective performance objectives. It is worth noting that in this formulation, it is assumed that the logic associated with constraint $C1$ is implemented in Algorithm \ref{pseudocode:Link_Avail}. 
\subsubsection{Knapsack Problem Formulation}
In the following section, we will show how this problem can be formulated as a 0-1 knapsack variant. The 0-1 knapsack problem can be formulated as follows:
%Put formulation here
\begin{equation}
    max \sum_{i=1}^N x_i \mathsf{cv}_i,
\end{equation}
\begin{align}
    \text{s.t.} \quad & \text{C1: } \sum_{i=1}^N \mathsf{wt}_i \leq W,
\end{align}
where $\mathsf{cv}_i$ is the value associated with an item, $\mathsf{wt}_i$ is the weight of each item, and $W$ is the maximum total weight of the knapsack. Indicating that the objective is to maximize the total value of the items placed in the knapsack while remaining below the total weight constraint. The problem definition described in the previous subsection can be reformulated in a manner similar to the 0-1 knapsack problem. Firstly, the weight constraint $W$ for our problem can be defined as the initial number of packets $\mathsf{DV}_{init}$. However, unlike the traditional 0-1 knapsack problem, this constraint is soft rather than hard. The items in this case are contacts in the downlink period. The value of each item is defined as
\begin{equation}
    \mathsf{cv} = 
\begin{cases}
  (\beta d_i - \upsilon e_i)& \text{if } d > 0, \\
    -\upsilon e_i& \text{if } d= 0,
\end{cases}
\end{equation}
where $\beta$ and $\upsilon$ are tuning parameters. The weight of each item is defined as
%Change this to be the possible packets
\begin{equation}
    \pgm{\mathsf{wt}=\frac{\mathsf{dr} \ \mathsf{CA} }{\mathsf{pl} \ \mathsf{lr}}.}
\end{equation}
To support the definition of our problem as a soft knapsack problem, the following simple function is defined:
%Define soft constraint here
\begin{equation}
    m(a,b)=\begin{cases}
        a \  \text{if } a \geq b,\\
        b \ \text {if } b>a.
    \end{cases}
\end{equation}
The overall objective is given as
\begin{equation}
    max \sum x_i \mathsf{cv}_i- (\beta + \upsilon)m(0,\sum x_i \mathsf{wt}_i-W),
\end{equation}
with the weight and value of each item stochastically determined by orbital dynamics and weather conditions. In addition, given that the contacts are only available in sequential order, the problem must be viewed as online, as we cannot return to old, unused contacts. Unlike in the previous formulation, the constraint $C1$ is not present explicitly.

The knapsack problem and its many variants have been widely studied. Knapsack problems are generally NP-hard, hence heuristics have traditionally been applied to achieve near-optimal results in polynomial time. However, as complexity and the number of constraints increase, developing effective heuristics becomes more challenging. Hence, using deep reinforcement learning (DRL) to tackle this problem has been proposed from a scalability perspective for larger knapsack variant problems \cite{KP_RL_Attention}.

In stochastic knapsack problems, adaptive strategies have been shown to outperform static strategies \cite{adaptive_stochastic}. 

To achieve these objectives, both static and adaptive schemes were developed; these are defined in the following sections.
\section{Downlink Scheduling Schemes}
\label{sec:proposed_schemes}
\pgm{In the following section, the downlink scheduling schemes and relevant baselines are defined. Notably, relevant downlink scheduling schemes are classified in terms of static schemes and adaptive schemes. Finally, the time complexity of each scheme is determined.}
\subsection{Static Techniques}
\paragraph*{Baseline Scheme}
As noted in Section \ref{sec:background}, CGR is a common routing algorithm for delay-tolerant networks, providing the ability to route in networks with disjoint paths effectively. Although CGR is designed to handle predictable link failure due to loss of LoS, it does not consider stochastic environmental events such as adverse weather conditions. Hence, in the context of a single satellite attempting to downlink information to the ground, CGR will use all available contacts until all data is delivered, as shown in
\begin{equation}
     x_n = 
\begin{cases}
  1& \text{if } \sum_{i=1}^n d_i < \mathsf{DV}_{init}, \\
    0& \text{otherwise},
\end{cases}
\end{equation}
where $x$ is the decision variable for utilizing a contact and $n$ is a certain contact in the downlink period. The main advantage of CGR is that it maximizes the delivery ratio; however, it does not consider energy efficiency.

%Baseline in line description

%Baseline pseudocode
%\begin{algorithm}
%\caption{Baseline Downlink Routing Scheme}
%\label{pseudocode:Baseline_CGR}
%\begin{algorithmic}[1]
%    \State $CVR \gets V$
%    \State $DV \gets \text{Number of Packets to deliver}$
%    \State $SGC \gets \text{Satellite to Ground Contacts}$
%    \State $A \gets 0$
%    \For{$i \gets 1 \ \mathbf{to} \ \text{Length}(SGC)$}        
%        \If{$DV>0$}
%            \State $d,a \gets LinkAvailability(SGC(i),DV)$
%            \State $A \gets A+a$
%            \State $DV \gets DV-d$
%        \EndIf
%    \EndFor
%    \State $Output \gets DV,A$
%\end{algorithmic}
%\end{algorithm}
\paragraph*{Threshold Schemes}
A simple way to improve the energy efficiency of basic CGR is by applying thresholds based on the cloud cover forecast. Given a cloud cover threshold value of $T$ and a current cloud cover percentage $CC$, the scheme logic can be described as follows:
\\
\begin{equation}
\text{Contact Decision}= 
\begin{cases} 
x_n=1& \text{if }  \mathsf{CC} <=  T \text{ and } \sum_{i=1}^{n} d < \mathsf{DV}_{init}\\
x_n=0 & \text {otherwise}\\
\end{cases}.
\end{equation}
%This logic is applied in Algorithm \ref{pseudocode:Threshold}, which outputs the same as Algorithm \ref{pseudocode:Baseline_CGR}. \\
%\begin{algorithm}
%\caption{Single Threshold  Routing Scheme}
%\label{pseudocode:Threshold}
%\begin{algorithmic}[1]
%    \State $\CVR \gets V$
%    \State $DV \gets \text{Number of Packets to deliver}$
%    \State $SGC \gets \text{Satellite to Ground Contacts}$
%    \State $T \gets \text{Selected Threshold}$
%    \State $A \gets 0$
%    \For{$i \gets 1 \ \mathbf{to} \ \text{Length}(SGC)$}          
%        \If{$DV>0 \ \mathbf{and} \ SGC(i)<T$}
%            \State $d,a \gets LinkAvailability(SGC(i),DV)$
%            \State $A \gets A+a$
%            \State $DV \gets DV-d$
%        \EndIf
%    \EndFor
%    \State $Output \gets DV,A$
%\end{algorithmic}
%\end{algorithm}

The effect of different cloud cover thresholds depends on the distribution of weather conditions. The obvious benefit of threshold schemes is their low computational cost, and in many cases, calculations can be done on the ground. This can be particularly beneficial for resource-constrained systems, such as CubeSats.

%Threshold tuning description
Although threshold schemes provide several aforementioned advantages, setting the correct threshold for a given ground station configuration is challenging. To aid in selecting cloud cover threshold values, Algorithm \ref{pseudocode:Thresh_Tune} was developed. This algorithm utilizes historical weather data and a set data volume distribution to determine a suitable threshold. It is assumed that this algorithm would be implemented before launching a satellite or before a change in the ground station configuration, with calculations conducted on the ground.
%Put pseudocode here
\begin{algorithm}
\caption{Threshold Tuning Algorithm}
\label{pseudocode:Thresh_Tune}
\begin{algorithmic}[1]
    \State $\mathsf{nTe} \gets \text{Number of Test Episodes For Each Setting}$
    \State $\mathsf{Tgr} \gets \text{Granularity of Threshold Step}$
    \State $\mathsf{nTv} \gets \text{Number of Threshold Values To Test}$
    \State $T \gets 100\%$
    \State $\mathsf{DRtol} \gets \text{Tolerance of Delivery Ratio Degradation}$
    
    \For{$j=1 \to \mathsf{nTv}$}        
        
            \For{$i=1 \to \mathsf{nTe}$}
                \State $\mathsf{Tsr},A \gets \text{ThresholdScheme}(\mathsf{CVR},\mathsf{DV},\mathsf{SGC},T)$
                \State $\mathsf{Bsr}(i),A \gets \text{BaselineScheme}(\mathsf{CVR},\mathsf{DV},\mathsf{SGC})$
            \EndFor
        \If{$\text{mean}(\mathsf{Tsr})/\text{mean}(\mathsf{Bsr})>1-\mathsf{DRtol}$}
            \State $T \gets T-\mathsf{Tgr}$
        \Else
        \State $Output\gets T$
        \EndIf
    \EndFor    
\end{algorithmic}
\end{algorithm}
This approach can be extended to use multiple thresholds by applying it for each range of data volume values covered by a separate threshold. The benefit of using multiple thresholds is that it yields a solution that is much more robust to changes in the data volume distribution than a single threshold.

\paragraph*{Sorting Algorithm}
Another approach is to sort contacts based on predicted cloud cover conditions. This scheme is defined in Algorithm \ref{pseudocode:Sort_Algo}. Given an initial number of packets to send, predicted weather conditions,  the lengths of future contacts,  \pgm{and volume margin value $\tau$}, this scheme outputs the selection of contacts to be used for downlinking the data. \pgm{In this case, $\tau$ is a continuous value that controls how aggressive the sorting algorithm is in prioritizing energy efficiency versus delivery ratio}. This results in a solution similar to the one presented in \cite{Lyu1}.
\begin{algorithm}
\caption{Static Sorting Algorithm}
\label{pseudocode:Sort_Algo}
\begin{algorithmic}[1]
    \State $\mathsf{DV} \gets \text{Initial Number of Packets To Send}$
    \State $\mathsf{SGC} \gets \text{Space-to-Ground Contacts}$
    \State $\mathsf{SC}\gets \text{Initialized To All Zeros}$
    \pgm{\State $\mathsf{\tau} \gets \text{Volume Margin}$}
     \State $E \gets 0$
    \For{$i \gets 1 \ \mathbf{to} \ \text{Length}(\mathsf{SGC})$}
        \State $\mathsf{SC}(i) \gets [\mathsf{SGC}(i,1),\mathsf{SGC}(i,2),i]$
    \EndFor
    \State $\mathsf{SC} \gets\text{Sort According To Cloud Cover}$
    \State $D \gets \mathsf{DV \pgm{\tau}}$
    \State $X \gets \text{Initialized To All Zeros}$
    \For{$i \gets 1 \ \mathbf{to} \ \text{Length}(\mathsf{SGC})$}        
        \If{$D>0 \ \mathbf{and} \ S(i,1)<100$}
           
            \State $X(\mathsf{SC}(i,3)) \gets 1$
            \State $D \gets D-(\mathsf{SC}(i,2)\mathsf{SC}(i,1)/100)$
        \EndIf
    \EndFor
    \For{$i \gets 1 \ \mathbf{to} \ \text{Length}(X)$}
        \If{$X(i)==1$}
            \State $d,e \gets \text{LinkAvailability}(\mathsf{SGC}(i),\mathsf{DV})$
            \State $E \gets E+e$
            \State $\mathsf{DV} \gets \mathsf{DV}-d$
        \EndIf
    \EndFor
    \State $Output \gets \mathsf{DV},E$
\end{algorithmic}
\end{algorithm}
%Discussion of some limitations
\pgm{Although this static sorting algorithm is less sensitive than the threshold downlink scheduling schemes, it still requires tuning of $\tau$ if there is substantial uncertainty associated with the link availability and predicted weather conditions. This coarse method set of parameter settings does not explicitly account for the stochastic nature of the link availability, as is the case with the adaptive techniques described in Section \ref{adapt_techs}.}
.
\subsection{Adaptive Techniques}
\label{adapt_techs}
\paragraph*{Adaptive Sorting}
%Describe adaptive sorting
To compensate for the stochastic nature of link availability, recalculating the contacts to be selected for use after transmission is highly desirable from a performance perspective. In Algorithm \ref{pseudocode:Adaptive_Sort_Algo}, similar to the Algorithm \ref{pseudocode:Sort_Algo}, the inputs consist of the overall number of packets to send, the weather conditions predicted to be present at each contact, the length of each contact \pgm{and the volume margin}.
\begin{algorithm}
\caption{Adaptive Sorting Algorithm}
\label{pseudocode:Adaptive_Sort_Algo}
\begin{algorithmic}[1]
    \State $\mathsf{DV} \gets \text{Initial Number of Packets To Send}$
    \State $\mathsf{SGC} \gets \text{Space-to-Ground Contacts}$
    
    \State $\mathsf{SC}\gets \text{Initialized To All Zeros}$
    \pgm{\State $\mathsf{\tau} \gets \text{Volume Margin}$}
    \For{$i \gets 1 \ \mathbf{to} \ \text{Length}(\mathsf{SGC})$}   
        \State $\mathsf{SC}(i) \gets [\mathsf{SGC}(i,1),\mathsf{SGC}(i,2),i]$
    \EndFor
    \State $\mathsf{SC} \gets\text{Sort According To Cloud Cover}$
    \State $D \gets \mathsf{DV\mathsf{\pgm{\tau}}}$
    \State $X \gets \text{Initialized To All Zeros}$
    \For{$i \gets 1 \ \mathbf{to} \ \text{Length}(\mathsf{SGC})$}          
        \If{$D>0 \ \mathbf{and} \   \mathsf{SC}(i,1)<100$}
                \State $X(\mathsf{SC}(i,3)) \gets 1$
                \State $D \gets D-(\mathsf{SC}(i,2) \mathsf{SC}(i,1)/100)$
        \EndIf
    \EndFor
    \For{$i \gets 1 \ \mathbf{to} \ \text{Length}(\mathsf{SGC})$}
            \If{$X(i)==1$}
                \State $d,e \gets \text{LinkAvailability}(\mathsf{SGC}(i),\mathsf{DV})$
                \State $E \gets E+e$
                \State $\mathsf{DV} \gets \mathsf{DV}-d$
                \State $\mathsf{SC}_2 \gets \mathsf{SGC}(i:\text{maxIndex}(\mathsf{SGC})) $
                \State $\mathsf{SC}_2 \gets\text{Sort According To Cloud Cover}$
                \State $ D_2 \gets \mathsf{DV \pgm{\tau}}$
                \For{$i \gets 1 \ \mathbf{to} \ \text{Length}(R)$}        
                     \If{$D_2>0 \ \mathbf{and} \   \mathsf{SC}_2(j,1)<100$}
                        
                            \State $X(\mathsf{SC}_2(j,3)) \gets 1$
                            \State $D_2 \gets D_2-\mathsf{SC}_2(j,2)$
                    
                    \EndIf
                \EndFor
            \EndIf
        \EndFor
    \State $Output \gets \mathsf{DV},E$
\end{algorithmic}
\end{algorithm}

\paragraph*{RL and DRL Agent Definition}
The sequential selection of contacts fits well with Markov decision processes\pgm{,} and hence, RL is an attractive approach. In \cite{fettes2025energyefficientsatelliteiotoptical}, the potential for using DRL to achieve the objectives discussed in Section \ref{sec:systemmodel} is explored. The formulation of the problem in terms of observation and action space is as follows. As ground stations are assumed to be far enough apart not to allow simultaneous communications, the action space is restricted to $\{0, 1\}$. In this case, when the agent selects action zero, the satellite does not attempt to send information to the ground, whereas when action one is selected, the satellite attempts to transmit to the ground, with success determined by Algorithm \ref{pseudocode:Link_Avail}.

It is assumed that the agent is located on the satellite node to provide access to data storage. The observation information available to the agent for each decision is  
\begin{equation}
 \begin{bmatrix}
\text{$\mathsf{CC} \quad$} 
\text{$\mathsf{DV} \quad$} 
\text{$\mathsf{CVR} \quad$} 
\text{$\mathsf{SGC}$}
\end{bmatrix}^T,\\
\end{equation}
where $\mathsf{CC}$ is the cloud cover for the next contact, $DV$ is the remaining data volume to be delivered, $\mathsf{CVR}$ is the remaining system capacity. $\mathsf{CVR}$ can be defined as follows:
%Insert Equation here
\begin{equation}
\mathsf{CVR}=V-\sum_{i=1}^{m} (v_i),
\end{equation}
 given $m$ is the current contact and $V$ is the initial total contact capacity defined as
 \begin{equation}
    V=\sum_{i=1}^{N} v_i.
\end{equation}
Finally, $SGC$ is a matrix containing all future weather and contact information for the downlink period. This matrix is an N by two matrix shown in the equation below:
%Put example matrix here
\begin{equation}
\mathsf{SGC}=
\begin{bmatrix}
\mathsf{CC}_1 & \mathsf{CC}_2 & \cdots & \mathsf{CC}_N \\
v_1 & v_2 & \cdots & v_N
\end{bmatrix}^T,
\end{equation}
with each matrix element representing the weather conditions and the length of a contact between the satellite and the ground. These are updated after each environment update, with contacts that are no longer available being set to 100\% cloud cover. For equal-length contacts, this matrix can be simplified to an N-by-1 matrix containing only the weather conditions.

A challenge with the environment defined above is the size of the state-action space when using conventional tabular reinforcement learning algorithms such as Q-learning. Although continuous observations can be discretized, this is an imperfect solution that reduces the granularity of observations and the agent's maximum achievable performance. 

%The size of the state-action space considering the simplifications discussed in Section \ref{sec:system_Arc} can be calculated using the following process. Firstly, in terms of $SGC$, the cloud cover can be a value between 0 and 1 with increments of 0.1. Hence, as there are 10 discrete values that this value can take in the simulated environment, 4 bits are required to represent this result accurately. Secondly, $\Omega$ is normalized to between 0 and 1 as well, with 100 potential discrete values, meaning that 7 bits are required to represent this observation parameter. Thirdly, $CVR$, given the same range as $\Omega$, also requires 7 bits to represent all the possible values. Finally, $SGC$ is a set of 10 values each between 0 and 1, with 10 possible values each, meaning that 4 bits are required for each of these values. Hence, the total size of the state-action space without any reduction in the quality of information provided to the agent is $2^{60}$ bits, accounting for the action space being binary. Therefore, for each observation field, the granularity of information represented in the Q-table is reduced to 2 bits (4 levels). This significantly reduces the size of the state-action space at the cost of the quality of the information provided to the agent. Hence, given this reduction in granularity, the total state-action space size is reduced to $2^{26}$, which remains very large nonetheless.

Given the large state-action space, deep reinforcement learning (DRL) was also tested, as it eliminates the need to reduce the quality of the information provided to the agent. Due to the discrete nature of the action space, double deep Q-Networks (DDQN) were assessed to be a suitable DRL algorithm. The observation information provided to the DDQN implementation was modelled as continuous values ranging from zero to one.

The objectives presented in Section \ref{sec:problem_form} are to maximize the delivery ratio and, where possible, improve energy efficiency. A suitable reward function must be defined for both the RL and DRL schemes. At each step, the agent chooses an action from the action space. If the agent decides not to utilize a contact, the agent receives zero reward. Otherwise, the agent receives a reward defined by the number of packets delivered during the current time step and the efficiency with which these packets are delivered. The reward will be positive. If the agent successfully delivers any packets during an environment step. The only circumstance that results in a negative reward for the agent is if a contact is selected and no packets are delivered. Improving energy efficiency is rewarded by allocating increased positive rewards for using more efficient contacts. The terminal reward is defined by the overall delivery ratio and energy efficiency. These components are outlined below:
\begin{equation}
\text{Total Rewards}=\text{Terminal Reward}+\sum(\text{Step Reward}).
\end{equation}
The delivery ratio and the total utilized contact time for an episode are defined as $DR$ and $CT$, respectively. With the scaling factor $c$, the terminal reward is defined as  

\begin{equation}
\text{ER}= 
\begin{cases} 
  2  c  \frac{\mathsf{DR}}{\mathsf{CT}}  & \text{if } \mathsf{DR} ==1,\\
c  \mathsf{DR} & \text{if } \mathsf{DR} <1. \\
\end{cases}
\end{equation}
The step reward ($\mathsf{SR}$) consists of components $F1$,$F2$, $F3$:
\begin{equation}
F1=\frac{c}{\mathsf{DV}_{init}} d_i,
\end{equation}
\begin{equation}
F2=\frac{c}{\mathsf{DV}_{init}} d_i\frac{e_i}{e_i+v_i},
\end{equation}
\begin{equation}
    F3=-e_i  \frac{c}{2 \mathsf{DV}_{init}},
\end{equation}
\begin{equation}
\mathsf{SR}= 
\begin{cases} 
F1-F2  & \text{if } F1 > 0 \text{ and Action=1}, \\
-F3 & \text{if } F1 = 0 \text{ and Action=1},\\
0 & \text{if } \text{Action}=0.\\
\end{cases}
\end{equation}
%Need to improve the reasoning behind the reward function
In the equation above, $\mathsf{DV}_{init}$ is the data volume that needs to be delivered during the transmission window, $d$ is the number of packets delivered during the current time step, $e$ is the number of unsuccessful transmissions during the current time step, and $v$ is the contact volume. The total reward for an episode is the sum of the total SR and ER values.
%\begin{equation}
%    R=\sum_{i=1}^m \mathsf{SR}_i+\mathsf{ER}.
%\end{equation}
%We can put the environment description here

Given the discrete nature of the action space, value-based RL and DRL algorithms were selected. The two algorithms selected were Q-learning and DDQN. For the aforementioned algorithms, the observation information provided to both algorithms must include the future contact information $SGC$, which must be updated at each step. This involves removing from the $SGC$ the contacts that are no longer available and reorganizing the matrix to reflect the future contacts. This is due to the online nature of the problem. In addition, for the Q-learning algorithm, due to the continuous nature of the observation space, discretization must be conducted. This reduces the granularity of the information provided to the agent, but it is necessary due to the tabular nature of the algorithm. Q-learning was selected primarily as a point of comparison for the DDQN implementation. This was to ensure the added complexity of DDQN is justified by increased performance when compared to simple tabular RL approaches.

The MDP environment logic was designed to incorporate the system dynamics defined in Algorithm \ref{pseudocode:Link_Avail} and the previously defined observation information. This environment is defined in Algorithm \ref{pseudocode:MDP_Env}.
\begin{algorithm}
\caption{Environment Logic}
\label{pseudocode:MDP_Env}
\begin{algorithmic}[1]
    \State $\mathsf{DV} \gets \text{Number of Packets to deliver}$
    \State $\mathsf{SGC} \gets \text{Satellite to Ground Contacts}$
    \State $\mathsf{TS} \gets \text{Max Timestep}$
    \State $E \gets 0$
    \State $k \gets 0$
   \While{$k<=\mathsf{TS} \space \ \& \ \space DV>0$}
        \State Observe current Observations
        \State Agent selects action according to trained policy $\pi$
        \If{$action==1$}
            \State $d_k,e_k \gets \text{LinkAvailability}(\mathsf{SGC}(k),\mathsf{DV})$
            \State $E \gets E+e$
            \State $D \gets D+d$
            \State $\mathsf{DV} \gets \mathsf{DV}-d$
            \State Obtain Step Reward $\mathsf{SR}_k$
        \Else
            \State Obtain Step Reward $\mathsf{SR}_k$
        \EndIf
        \State Format observation information
        \State $k \gets k+1$
    \EndWhile
    \State Obtain terminal reward $\mathsf{ER}$
    \State $R \gets \sum_{i=1}^k \mathsf{SR}_i+\mathsf{ER}$
    \State $Output \gets \mathsf{DV},E,D,R$
\end{algorithmic}
\end{algorithm}

\subsection{Time Complexity Analysis}
%Put the Big-O notation
The worst-case theoretical time complexity performance of the algorithms presented in Section \ref{sec:proposed_schemes} is highly relevant from a scalability perspective. As operations conducted on the satellite component of the network are the most sensitive to scalability issues, the timing complexity analysis was limited to these operations. This sensitivity is due to the resource-limited nature of satellite hardware. Hence, the time complexity of training the proposed DDQN solution and tuning the threshold algorithms, as specified in Algorithm \ref{pseudocode:Thresh_Tune}, was not considered.
\begin{table}[ht]
\centering
\caption{List of Time Complexity Symbols}
\label{tab:symbols_TS}
\begin{tabular}{ll}
\hline
\textbf{Symbol} & \textbf{Description} \\
\hline
$c_{\mathsf{LA}}$& Link Availability Computations\\
$c_{\mathsf{SA}}$ & SC Configuration Computations \\
$c_{\mathsf{SSO}}$ & Contact Volume Estimation Computations\\
$y$ & Neural Network Output \\
$x_{\mathsf{nn}}$ & Neural Network Inputs\\
$w_{\mathsf{nn}}$ & Neural Network Weights\\
$b$ & Neural Network Bias\\
$h$ & ReLu Activation Function \\
$I$ & Number of Inputs\\
$nn$ & Number of Neurons\\
$c_{\mathsf{FP}}$ & Forward Propagation Computations\\
$c_{\mathsf{obs1}}$ & Observation Info Formatting Computations\\
$c_{\mathsf{obs2}}$ & Observation Info Formatting Computations\\
$c_{a}$ & Action Selection Computations\\
$c_{\mathsf{obs3}}$ & Observation Info Discretization Computations\\
\hline
\end{tabular}
\end{table}
The time complexity of each algorithm is defined in terms of the number of contacts $N$ in each downlink period. For CGR and threshold schemes, the relationship between time complexity and $N$ is linear under worst-case assumptions. In addition, the constant $c_{LA}$ incorporates the time complexity of the link availability calculations, resulting in the following expressions:
\begin{equation}
    \text{CGR Timing Complexity}=O(N) c_{\mathsf{LA}},
\end{equation}
\begin{equation}
    \text{Threshold Timing Complexity}=O(N) c_{\mathsf{LA}}.
\end{equation}
In terms of Algorithm \ref{pseudocode:Sort_Algo}, in addition to the linear relationship between $N$ and the number of times all the processes must be run, the sorting algorithm must sort the available contacts using a standardized sorting algorithm (assumed to be merge sort for this analysis). The complexity of sorting is defined as
\begin{equation}
    \text{Sorting Complexity}=O(N\text{log}N).
\end{equation}
While the complexity of other required operations within the static sorting algorithms is defined as 
\begin{equation}
    \text{Auxiliary Complexity}=O(N)( c_{\mathsf{SA}}+c_{\mathsf{SSO}}+ c_{\mathsf{LA}}),
\end{equation}
where $c_{\mathsf{SSO}}$ and $c_{\mathsf{SA}}$ are the operations associated with estimating the number of contacts that must be used and configuring the matrix $\mathsf{SC}$. These results can be simplified to an overall time complexity 
\begin{equation}
    \text{Static Sorting Time Complexity}=O(N\text{log}N).
\end{equation}
 Unlike Algorithm \ref{pseudocode:Sort_Algo}, Algorithm \ref{pseudocode:Adaptive_Sort_Algo} must conduct sorting operations every time a contact is used. In the worst-case scenario, this would entail conducting sorting operations for every contact $N$. Hence, the overall time complexity can be defined as follows:
\begin{equation}
    \text{Sort Complexity}=O({N}^2\text{log}N),
\end{equation}
    
\begin{align}
    \text{Auxiliary Complexity}=O(N^2) c_{\mathsf{SSO}} +O(N) c_{\mathsf{SA}}\\+O(N) c_{\mathsf{LA}} +O(N) c_{\mathsf{SSO}},
\end{align}
\begin{equation}
    \text{Adaptive Sorting Time Complexity}=O({N}^2\text{log}N).
\end{equation}

The time complexity of RL- and DRL-based approaches can be determined by analyzing several algorithm components. Firstly, the time complexity of selecting an action for each environment step must be examined. Secondly, as the observation information provided to the RL and DRL algorithms must be formatted, these operations must be taken into account when evaluating the overall time complexity. Finally, operations associated with forward propagation through the neural network of the DRL solution must be considered.

Forward propagation for a single neuron is defined as
%Put equation with explanation here
\begin{equation}
    y=h(x_{\mathsf{nn}}w_{\mathsf{nn}}^T+b),
\end{equation}
where $h(y)$ is the activation function. In this case, the ReLu function is used and is defined as follows:
\begin{equation}
    h(y)=
    \begin{cases}
       y,\text{ if } y=<0,\\
        0,\text{ if } y>0.
    \end{cases}
\end{equation}
The additional variable are $x_{\mathsf{nn}}$, the input values, $w_{\mathsf{nn}}$, the weights, and $b$, the bias. In a large neural network layer comprising many neurons, these operations are often vectorized. As each neural network layer is fully connected, the number of operations that must be conducted to obtain the output for each neuron is equal to
\begin{equation}
    \text{Number of Operations at Each Neuron}=I  c_{\mathsf{FP}},
\end{equation}
where $c_{\mathsf{FP}}$ is the number of calculations required for a single input and $I$ is the number of inputs. The number of contacts in the downlink period only affects the number of operations in an input neural network layer.
%\begin{figure}%[H]
%\centering
%\includegraphics[width=0.98\linewidth]{fig/Comparison_Baseline_Schemes/DQN_Agent_Diagram.pdf} 
%  \caption{Neural network structure of the implemented DQN agent.}
%  \label{fig:DQN_struct}
%\end{figure}
The inference time complexity is linearly proportional to the number of contacts in a downlink period, and hence can be defined as 
\begin{align}
    \text{Inference Complexity}=O(N) \mathsf{nn} c_{\mathsf{FP}}+\\3 c_{\mathsf{FP}} \mathsf{nn}+3c_{\mathsf{FP}}\mathsf{nn}+Nc_{\mathsf{FP}}\mathsf{nn}  +2 c_{\mathsf{FP}},
\end{align}
where $\mathsf{nn}$ is the number of neurons in the neural networks. In terms of the time complexity of the observation information formatting, this was found to be
%Put equation here
\begin{equation}
    \text{Environment Complexity}= O(N)c_{\mathsf{obs1}}+O(N)c_{\mathsf{obs2}},
\end{equation}
where $c_{\mathsf{obs1}}$ and $c_{\mathsf{obs2}}$ capture observation info formatting operations that must be done at each environment step. 
%Finally selecting an action results in the following complexity:
%Put equation here
%\begin{equation}
%    \text{Action Selection}=O(C)
%\end{equation}
Making the combined time complexity for the DDQN scheme is
\begin{equation}
    \text{Total DDQN Inference Time Complexity}=O(N^2).
\end{equation}
%Next we must consider the time complexity of Q-Learning
The time complexity of Q-learning can be calculated using a similar process to that of DDQN. Firstly, we know that selecting an action requires finding the maximum Q-value in the Q-table. The time complexity of this operation is directly proportional to the size of the action space $c_a$, hence it can be defined as 
\begin{equation}
    \text{Action Selection}=O(c_a).
\end{equation}
As the observation information must be discretized to ensure that the Q-table size does not exceed the memory available, several operations must be conducted to modify the observation information. The complexity of these operations is captured in 
\begin{equation}
    \text{Environment Complexity}=O(N) (c_{\mathsf{obs3}} + c_{\mathsf{obs1}}+ c_{\mathsf{obs2}}),
\end{equation}
where $c_{\mathsf{obs3}}$ is an additional observation formatting operation associated with discretizing the relevant information. All of these processes must be conducted in every environment step; hence, in a testing scenario, the overall time complexity is
\begin{equation}
    \text{Q-learning Total Time Complexity}=O(N^2).
\end{equation}
The results of the time complexity analysis show that the threshold and CGR schemes have the lowest complexity, while Algorithm \ref{pseudocode:Adaptive_Sort_Algo} is the least scalable. %However, in practical testing shown in Figure \ref{fig:Exec_time} for short downlink periods, both Q-learning and DQN were significantly slower than the other algorithms.

%Brief discussion and notes regarding Big O-Notation
\begin{table}[h]

\label{Time_Complex_Table}
\caption{Time complexity results for all algorithms.}
\centering
\renewcommand{\arraystretch}{1.3} 
\begin{tabular}{|c|c|}
\hline
\bigtablefont Algorithms & \bigtablefont Time Complexity \\
\hline
\bigtablefont Baseline CGR& \bigtablefont $ O(N) $\\
\hline
\bigtablefont Single Threshold Algorithm& \bigtablefont $O(N)$\\
\hline
\bigtablefont Multi-Threshold Algorithm& \bigtablefont $O(N)$\\
\hline
\bigtablefont Static Sorting& \bigtablefont $O(N\text{log}N)$\\
\hline
\bigtablefont Adaptive Sorting& \bigtablefont $O(N^2\text{log}N)$\\
\hline
\bigtablefont DDQN& \bigtablefont $O(N^2)$\\
\hline
\bigtablefont Q-learning &\bigtablefont $O(N^2)$\\
\hline
\end{tabular}
\end{table}

%As one can observe from the results in Figure \ref{fig:Exec_time}, the adaptive schemes, particularly the DQN scheme are significantly more computational intensive than the static approaches. This indicates a key cost-benefit trade-off with utilizing higher performing schemes.\\

%It is expected that increasing the length of the downlink period would result in an increase computation time for all schemes. In particular, Algorithm \ref{pseudocode:Adaptive_Sort_Algo} must sort the contacts frequently, hence testing was conducted to determine the impact of increased downlink period on this specific scheme.

%To determine the impact on computation time of increasing the downlink period, the downlink period was varied between 10, 20, 30 and 40 contacts in length. This indicates that Algorithm \ref{pseudocode:Adaptive_Sort_Algo} continues to show better computation time performance than DQN for a broad range of downlink period configurations.

\section{Simulation Settings}
\label{sec:Sim_settings}
\pgm{The simulation configurations, including the associated training and tuning configurations, are described in the section below. This includes the hyperparameter tuning approach for the RL and DRL algorithms, the tuning approach for threshold and sorting algorithms, and the specific simulation configurations that are used for performance evaluation in Section \ref{sec:evaluation}. The simulations are divided into two categories: general simulations and a case study.}

\subsection{General Simulation Configurations}
\label{Simulation_Config}
%General Configurations
\pgm{For the two general simulations, several key simulation parameters were set to be constant. The number of contacts $N$ in each downlink period was set to 10, and the volume of each contact $i$ was set to be of equal length. This more simplistic simulation configuration allows for an in-depth analysis of each scheme in a highly controlled environment.}

\subsubsection{Uniform Data Volume Configuration}
\label{Data_vol_Config}
%Uniform data distribution
To assess the overall performance of all proposed schemes, a standardized configuration was developed with a wide range of data volumes. The data volume distribution was set to be uniformly distributed between 5\% and 100\% of the downlink period capacity without cloud cover (5 to 100 packets to send). This was the same configuration that was used for training and tuning the proposed schemes, providing a key benchmark for performance under normal operating conditions.
%Put equation here
%\begin{equation}
%    \text{Data Volume}=Uniform(5,100)
%\end{equation}

\begin{figure*}[htbp]
    \centering
    % First row: two subfigures side by side
    \begin{minipage}[b]{0.4\textwidth}
        \centering
        \includegraphics[width=\linewidth]{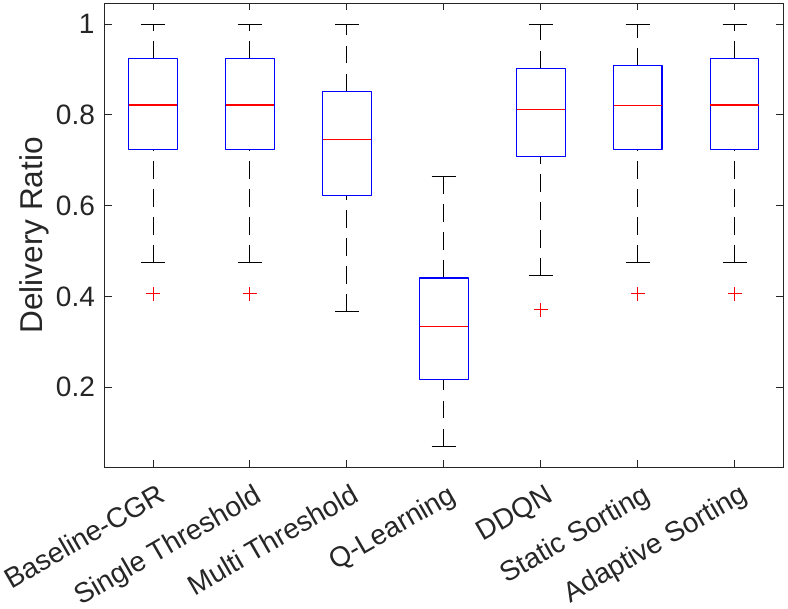}
        \subcaption{Delivery ratio.}
        \label{fig:Uniform_Delivery}
    \end{minipage}
    \hspace{0.5cm} % space between the two columns
    \begin{minipage}[b]{0.4\textwidth}
        \centering
        \includegraphics[width=\linewidth]{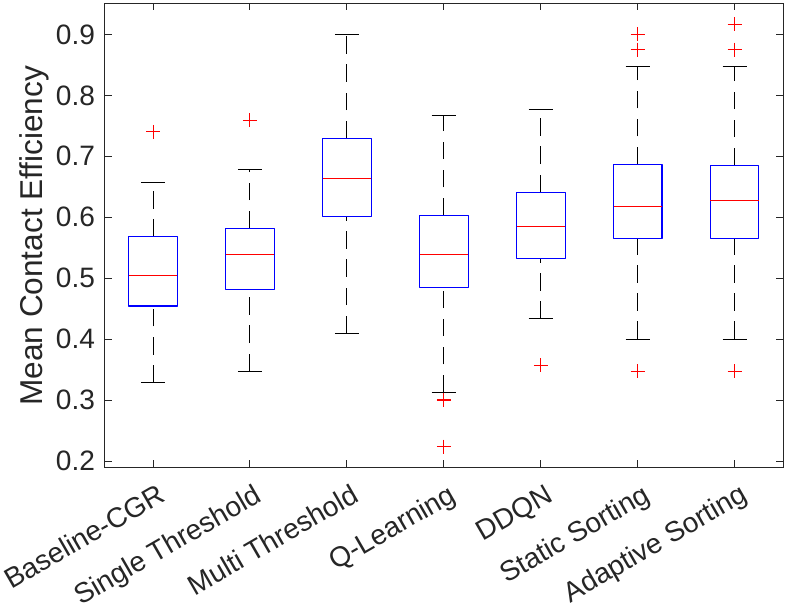}
        \subcaption{Mean contact efficiency.}
        \label{fig:Uniform_Energy}
    \end{minipage}

    \vskip\baselineskip % vertical space between rows
    \caption{Results of schemes under uniformly distributed data volume and weather conditions.}
    \label{fig:Uniform_results}
\end{figure*}
\subsubsection{Variable Cloud Cover and Data Volume Configuration}
\label{variation_sim_settings}
%Variable Configuration
To evaluate the performance of the downlink scheduling schemes under more dynamic conditions, a second set of simulations was conducted, incorporating changes to both the data volume and cloud cover distributions. This was intended to simulate unexpected changes to ground station infrastructure and mission parameters. Therefore, no retraining or retuning was conducted for these configurations, as it is assumed that these might not be possible in many relevant scenarios. Three different cloud cover distributions were tested in combination with three different data volume distributions. Due to space limitations, the performance results of four key configurations are displayed in Section \ref{sec:evaluation}. 

These configurations were selected to demonstrate key performance when data volume is low or moderate across a wide range of cloud cover conditions. Generally, configurations with very high data volume values are not particularly interesting, as they result in all schemes reverting to strategies nearly identical to the baseline CGR. The specific simulation configurations are described in Table \ref{tab:config_table}. Unlike the configuration described in Section \ref{Data_vol_Config}, the cloud cover conditions for these experiments are normally distributed around the mean value, with a standard deviation of 20\%. 
%Put configurations here
\renewcommand{\arraystretch}{1.5}
\begin{table}[h]
\caption{Data volume and cloud cover distribution configurations.}
\centering
\begin{tabular}{|c|c|}
\hline
\bigtablefont Data Volume Value & \bigtablefont Cloud Cover Mean \\
\hline
\bigtablefont 0.1$V$& \bigtablefont 80\% \\
\hline
\bigtablefont 0.5$V$& \bigtablefont 80\% \\
\hline
\bigtablefont 0.5$V$& \bigtablefont 50\% \\
\hline
\bigtablefont 0.5$V$& \bigtablefont 30\%  \\
\hline
\end{tabular}

\label{tab:config_table}
\end{table}

%Timing simulations
%\subsubsection{Timing Test Configuration}
%A critical aspect of any downlink scheduling technique is the computational requirements of the algorithms. This is due to the limited computational capacity of satellites. 

%Both theoretical time complexity analysis and practical time complexity experiments were conducted.
%The practical testing was conducted in MATLAB 2024a, using a single thread and an Intel® Core™ Ultra 7 processor, 4.8 GHz. In terms of the data volume and cloud cover distribution configuration, these were the same as defined in Section \ref{Data_vol_Config}. Two tests were conducted, a general test for all developed schemes and a specific test analyzing the impact of increased downlink period length on  Algorithm \ref{pseudocode:Adaptive_Sort_Algo}.
\subsection{\pgm{Case Study Configuration}}
\label{sec:Case_Study_Config}
\pgm{To evaluate the performance of the downlink scheduling schemes in a more realistic scenario, a case study involving a single satellite transmitting data to optical ground stations located in several Canadian cities was developed.}
%Orbital Dynamics and settings
\pgm{Similar to the previous simulation configurations, the downlink period was limited to 10 contacts. To determine the length of the contacts simulations using MATLAB satellite communications toolbox were used to find the access intervals (a minimum elevation angle of 20 degrees was assumed) between a satellite in polar orbit (orbital parameters are specified in Table \ref{tab:Orbit_table}) and either Calgary and Inuvik for training or Ottawa and Calgary for training in 2023 and 2024. These series of access intervals are randomly selected during training and testing from the training or test years.}
\begin{table}[h]
\caption{\pgm{Orbital simulation parameters.}}
\centering
\begin{tabular}{|c|c|}
\hline
\bigtablefont \pgm{Orbital Parameters} & \bigtablefont \pgm{Values} \\ \hline
\bigtablefont \pgm{Altitude} & \bigtablefont \pgm{500km} \\ \hline
\bigtablefont \pgm{Eccentricity}& \bigtablefont \pgm{0} \\ \hline
\bigtablefont \pgm{Inclination}& \bigtablefont \pgm{$99.5^{\circ}$} \\ \hline
\bigtablefont \pgm{RAAN}& \bigtablefont \pgm{$0^{\circ}$} \\ \hline
\bigtablefont \pgm{True Anomaly}& \bigtablefont \pgm{$0^{\circ}$} \\ \hline
\bigtablefont \pgm{Argument of the Periapsis} & \bigtablefont \pgm{$0^{\circ}$} \\ \hline 
\end{tabular}
\label{tab:Orbit_table}
\end{table}
%Weather
\pgm{To obtain realistic cloud cover values for the ground station sites used in this simulation, historical weather data was obtained from \cite{meteoblue}.}
%Data Volume
\pgm{The data volume distribution for these simulations for both training and testing is identical relative to the total access time to the configuration described in Sections \ref{Data_vol_Config} and \ref{variation_sim_settings}.}
%Training and testing configuration

\pgm{In terms of link availability and weather modelling, the same model as defined in Algorithm \ref{pseudocode:Link_Avail} was used. Unlike the previous two simulation configurations, uncertainty was added to the observed weather data. This was modelled as a normally distributed value around the true cloud cover fraction, with a standard deviation of 0.2, for both training and testing.}
\subsection{RL and DRL Hyperparameter and Training Configuration}
The hyperparameters and training configurations of RL algorithms can affect their performance. Firstly, for \pgm{the general} simulation configurations, the DDQN and Q-learning were trained on cloud cover data uniformly distributed between 0 and 1 and data volume values between 5 and 100 percent of the total contact volume $V$. \pgm{While the case study DDQN agents were trained on uniformly distributed data volume values and historical weather data from 2023, using a ground station site configuration of Inuvik and Calgary.}

The hyperparameters used for the DDQN scheme are displayed in Table \ref{hyper_table}. \pgm{The general simulation configuration was} obtained via random search over 100 different hyperparameter configurations. \pgm{ While the case study hyperparameter tuning configuration was left as future work and kept as the default values for the MATLAB DDQN agent implementation}. For the Q-learning implementation, the hyperparameters were tuned using a simple grid search approach, as there were fewer hyperparameters to adjust. The resulting set of hyperparameters is displayed in Table \ref{Q_table}.
\begin{table*}
\renewcommand{\arraystretch}{1.5} 
\centering
\caption{Hyperparameter configuration.}
\begin{tabular}{|c|c|c|}
\hline
 \bigtablefont Hyperparameter & \bigtablefont General Simulations & \bigtablefont \pgm{Case Study}\\
\hline
 \bigtablefont $\alpha$ &  \bigtablefont 0.001782 & \bigtablefont \pgm{0.01} \\
\hline
 \bigtablefont Experience Buffer Size &  \bigtablefont 8656429 & \bigtablefont \pgm{10000} \\
\hline
 \bigtablefont $\gamma$ &  \bigtablefont 0.99 & \bigtablefont \pgm{0.99} \\
\hline
 \bigtablefont Epsilon Decay &  \bigtablefont 6.37e-6 & \bigtablefont \pgm{0.005}\\
\hline
 \bigtablefont MiniBatch Size &  \bigtablefont 75 & \bigtablefont \pgm{64}\\
\hline
 \bigtablefont Target Update Frequency &  \bigtablefont 1 & \bigtablefont \pgm{1}\\
\hline
 \bigtablefont L2 Regularization &  \bigtablefont 1e-9 & \bigtablefont \pgm{1e-4}\\
\hline
 \bigtablefont Target Smooth Factor &  \bigtablefont 0.02028 & \bigtablefont \pgm{1e-3}\\
\hline
 \bigtablefont Initial Epsilon Value &  \bigtablefont 1.0 & \bigtablefont \pgm{1.0}\\
\hline
 \bigtablefont Minimum Epsilon Value &  \bigtablefont 0.01 & \bigtablefont \pgm{0.01}\\
\hline
 \bigtablefont Number of Lookahead steps &  \bigtablefont 1 & \bigtablefont \pgm{1} \\
\hline
\end{tabular}
\label{hyper_table}
\end{table*}

\begin{table}
\renewcommand{\arraystretch}{1.5}
\centering
\caption{Q-learning hyperparameter configuration.}
\begin{tabular}{|c|c|}
\hline
\bigtablefont Hyperparameter & \bigtablefont Value\\
\hline
\bigtablefont $\alpha$ & \bigtablefont 0.00001  \\
\hline
\bigtablefont Epsilon Decay & \bigtablefont 0.00001\\
\hline
\bigtablefont Initial Epsilon Value & \bigtablefont 1.0\\
\hline
\bigtablefont Minimum Epsilon Value & \bigtablefont 0.005\\
\hline
\end{tabular}
\label{Q_table}
\end{table}
%Describe Q-Learning Observation info description
As we discussed in Section \ref{sec:proposed_schemes}, the observation information being provided to the Q-learning agent needs to be discretized, given the continuous nature of the observation space. 
\subsection{Threshold and Sorting Scheme Configuration}
The single-threshold and multi-threshold schemes were tuned using the same data volume and cloud cover distribution parameters as the RL schemes (with a different set of seeds than the simulations themselves), as shown in Algorithm \ref{pseudocode:Thresh_Tune}. The multi-threshold scheme was configured with five distinct data volume ranges (five thresholds). The thresholds configured for all simulations are displayed in Table \ref{tab:Thresholds}.

\begin{table}[]
    \renewcommand{\arraystretch}{1.5}
    \centering
     \caption{Threshold scheme configurations for general simulations.}
    \begin{tabular}{|c|c|c|}
        \hline
        \bigtablefont Data Volume Ranges& \bigtablefont Single &\bigtablefont  Multi \\
        \hline
        \bigtablefont  0-0.2$V$& \bigtablefont 0.9& \bigtablefont 0.1\\
         \hline
         \bigtablefont 0.2-0.4$V$&\bigtablefont 0.9&\bigtablefont 0.4\\
         \hline
         \bigtablefont 0.4-0.6$V$&\bigtablefont 0.9&\bigtablefont 0.8\\
         \hline
         \bigtablefont 0.6-0.8$V$&\bigtablefont 0.9&\bigtablefont 0.8\\
         \hline
         \bigtablefont 0.8-1.0$V$&\bigtablefont 0.9&\bigtablefont 0.8\\
         \hline
    \end{tabular}
   
    \label{tab:Thresholds}
\end{table}

\pgm{In terms of the sorting schemes, for the first two general simulations, $\tau$ was set to 1. But, for the case study, due to the weather forecast uncertainty, tuning was conducted, and for static sorting, $\tau$ was set to 1.3, while for adaptive sorting, the optimal value remained at 1. While the threshold scheme configurations for the case study are shown in Table \ref{tab:Thresholds_Case_study}.}
\begin{table}[]
    \renewcommand{\arraystretch}{1.5}
    \centering
     \caption{\pgm{Threshold scheme configurations for the case study.}}
    \begin{tabular}{|c|c|c|}
        \hline
        \bigtablefont \pgm{Data Volume Ranges}& \bigtablefont \pgm{Single} &\bigtablefont  \pgm{Multi} \\
        \hline
        \bigtablefont  \pgm{0-0.2$V$}& \bigtablefont \pgm{0.94}& \bigtablefont \pgm{0.78}\\
         \hline
         \bigtablefont \pgm{0.2-0.4$V$}&\bigtablefont \pgm{0.94}&\bigtablefont \pgm{0.92}\\
         \hline
         \bigtablefont \pgm{0.4-0.6$V$}&\bigtablefont \pgm{0.94}&\bigtablefont \pgm{0.99}\\
         \hline
         \bigtablefont \pgm{0.6-0.8$V$}&\bigtablefont \pgm{0.94}&\bigtablefont \pgm{0.99}\\
         \hline
         \bigtablefont \pgm{0.8-1.0$V$}&\bigtablefont \pgm{0.94}&\bigtablefont \pgm{0.99}\\
         \hline
    \end{tabular}
   
    \label{tab:Thresholds_Case_study}
\end{table}

%Hyperparameters, threshold tuning and sorting algorithm tuning

%Simulation setup, Training for DQN
%Inuvik Calgary Training
%Ottawa Calgary Testing
%Hyperparameter configuration
%Orbital Parameters
%Weather Data

\section{Performance Evaluation}
\label{sec:evaluation}
\pgm{In the section below, the performance of the schemes defined in Section \ref{sec:proposed_schemes} is analyzed in several different simulation configurations. The pros and cons of each scheme under various simulation conditions are discussed, along with potential real-world considerations when selecting a suitable downlink scheduling approach.}
\subsection{Uniform Data Volume Distribution}
\begin{figure*}[htbp]
    \centering
    % First row: two subfigures side by side
    \begin{minipage}[b]{0.45\textwidth}
        \centering
        \includegraphics[width=\linewidth]{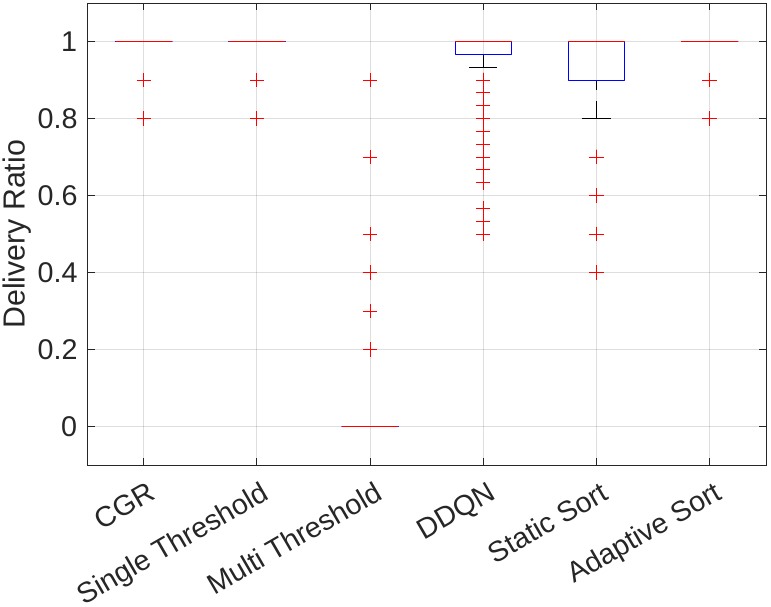}
        \subcaption{Delivery ratio when data volume is 10\% of capacity and mean cloud cover is 80\%.}
        %\subcaption{Delivery ratio-standard deviation 0 packets}
        \label{fig:low_packet_del}
    \end{minipage}
    \hspace{0.5cm} % space between the two columns
    \begin{minipage}[b]{0.45\textwidth}
        \centering
        \includegraphics[width=\linewidth]{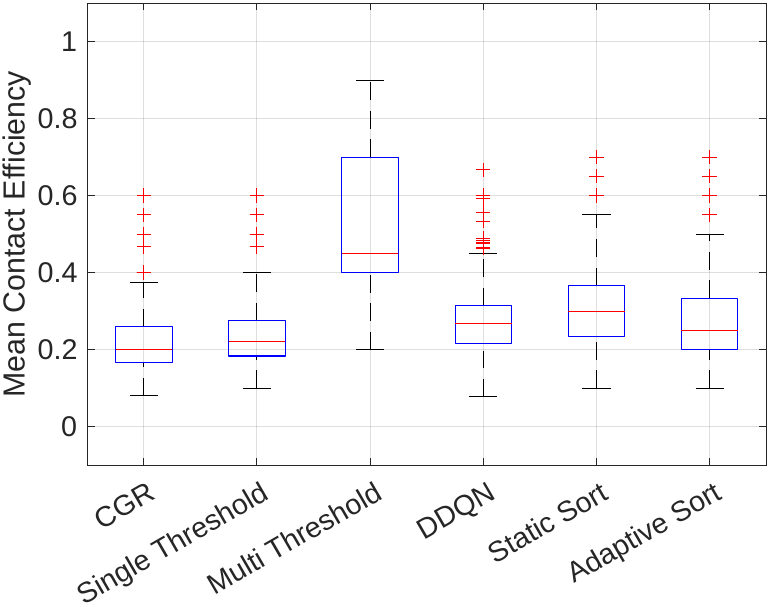}
        \subcaption{Energy efficiency when data volume is 10\% of capacity and mean cloud cover is 80\%.}
        %\subcaption{Mean contact efficiency-standard deviation 0 packets}
        \label{fig:low_packet_eff}
    \end{minipage}

    \vskip\baselineskip % vertical space between rows

    % Second row: two subfigures side by side
    \begin{minipage}[b]{0.45\textwidth}
        \centering
        \includegraphics[width=\linewidth]{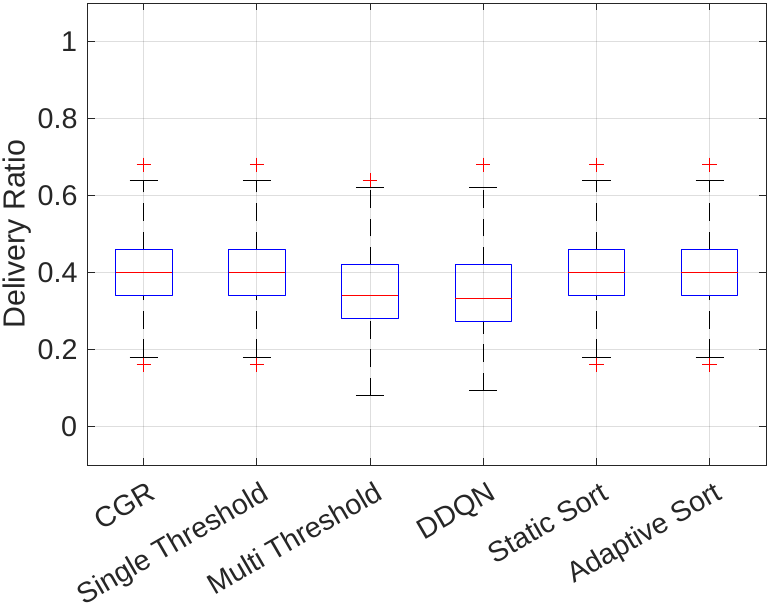}
        \subcaption{Delivery ratio when data volume is 50\% of capacity and mean cloud cover is 80\%.}
        %\subcaption{Delivery ratio-standard deviation 30 packets}
        \label{fig:high_packet_del}
    \end{minipage}
    \hspace{0.5cm} % space between the two columns
    \begin{minipage}[b]{0.45\textwidth}
        \centering
        \includegraphics[width=\linewidth]{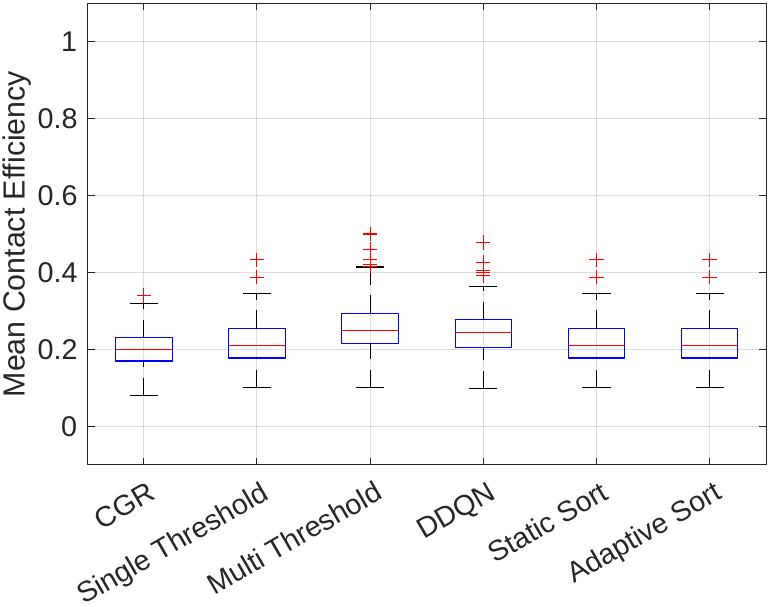}
        \subcaption{Energy efficiency when data volume is 50\% of capacity and mean cloud cover is 80\%.}
        %\subcaption{Mean contact efficiency-standard deviation 30 packets}
        \label{fig:high_packet_eff}
    \end{minipage}
    \caption{Results when cloud cover is high.}
    \label{fig:Packet_Tests}
\end{figure*}

Using three sets of 100 trials and the simulation settings defined in Section \ref{Simulation_Config}, the results shown in Figure \ref{fig:Uniform_results} were produced. From these results, several general observations can be made. Firstly, Q-learning is clearly an outlier, with poor delivery ratio performance and mediocre energy efficiency. This demonstrates that Q-learning, in particular, is poorly suited for tackling this problem; therefore, further testing was not conducted in subsequent sections. Secondly, one can observe that the multi-threshold scheme, although having the highest average mean contact energy efficiency with a 30\% improvement over baseline CGR, has the second-worst delivery ratio performance. This is not in line with the problem objectives, where the delivery ratio is given higher importance than energy efficiency. Thirdly, of the remaining schemes, the DDQN,  Algorithm \ref{pseudocode:Sort_Algo} and Algorithm \ref{pseudocode:Adaptive_Sort_Algo} improve average mean contact energy efficiency by between 14.8\% for DDQN to 24.7\% for Algorithm \ref{pseudocode:Adaptive_Sort_Algo} when compared to CGR. Hence, DDQN and the sorting algorithms show better energy efficiency performance than single threshold and baseline CGR, while only reducing mean delivery ratio by a maximum of 1.2\% for DDQN and less than 0.1\% for Algorithm \ref{pseudocode:Adaptive_Sort_Algo} relative to CGR. Hence, for this particular simulation configuration, DDQN, static sorting, and adaptive sorting are the best-performing schemes.

 However, these experiments only demonstrate performance for a single data volume and cloud cover distribution. In reality, it would be expected for any scheme to perform well under various configurations due to the site-specific nature of cloud cover distributions and the mission-dependent (or possibly customer-dependent) nature of data volume distribution. In addition, due to failures or changes in mission objectives, ground station sites or data volume distribution may change during a mission, requiring any energy saving scheme to be able to adapt without reconfiguration. In the following section, all the tested schemes are evaluated, assuming no retraining or retuning, to determine their resiliency to changes in weather and data volume.

\begin{figure*}[htbp]
    \centering
    % First row: two subfigures side by side
    \begin{minipage}[b]{0.45\textwidth}
        \centering
        \includegraphics[width=\linewidth]{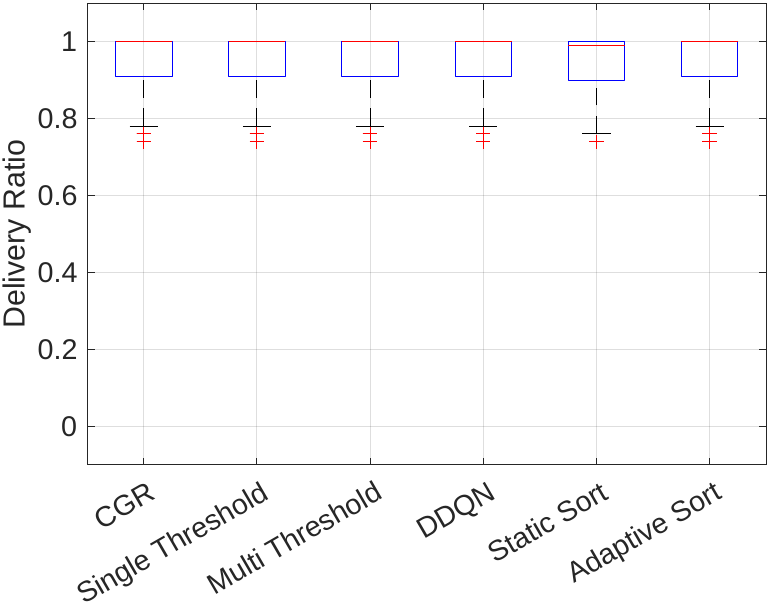}
        \subcaption{Delivery ratio when data volume is 50\% of capacity and cloud cover mean is 50\%.}
        %\subcaption{Delivery ratio-standard deviation 0 packets}
        \label{fig:DR_}
    \end{minipage}
    \hspace{0.5cm} % space between the two columns
    \begin{minipage}[b]{0.45\textwidth}
        \centering
        \includegraphics[width=\linewidth]{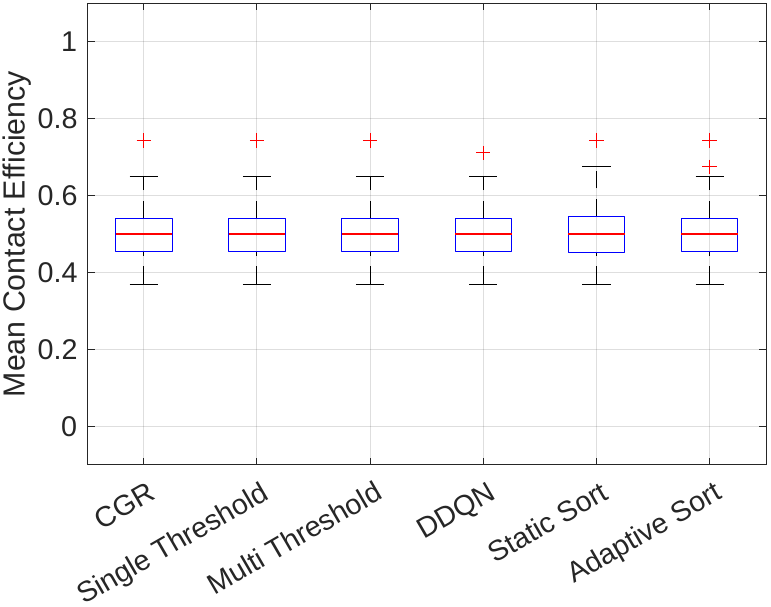}
        \subcaption{Energy efficiency when data volume is 50\% of capacity and mean cloud cover is 50\%.}
        %\subcaption{Mean contact efficiency-standard deviation 0 packets}
        \label{fig:EE_Medium_Data_Volume_Medium_Cloud_Cover}
    \end{minipage}

    \vskip\baselineskip % vertical space between rows

    % Second row: two subfigures side by side
    \begin{minipage}[b]{0.45\textwidth}
        \centering
        \includegraphics[width=\linewidth]{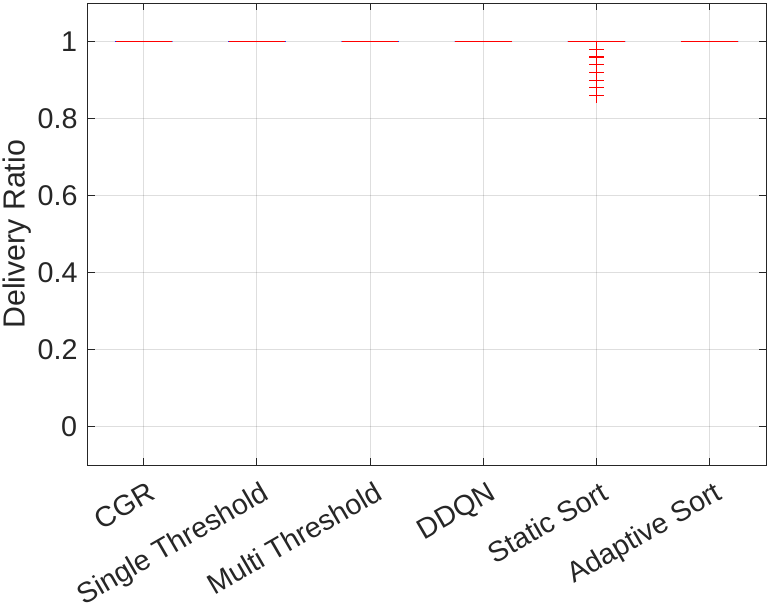}
        \subcaption{Delivery ratio when data volume is 50\% of capacity and mean cloud cover is 30\%.}
        %\subcaption{Delivery ratio-standard deviation 30 packets}
        \label{fig:DR_Medium_Data_Volume_Low_Cloud_cover}
    \end{minipage}
    \hspace{0.5cm} % space between the two columns
    \begin{minipage}[b]{0.45\textwidth}
        \centering
        \includegraphics[width=\linewidth]{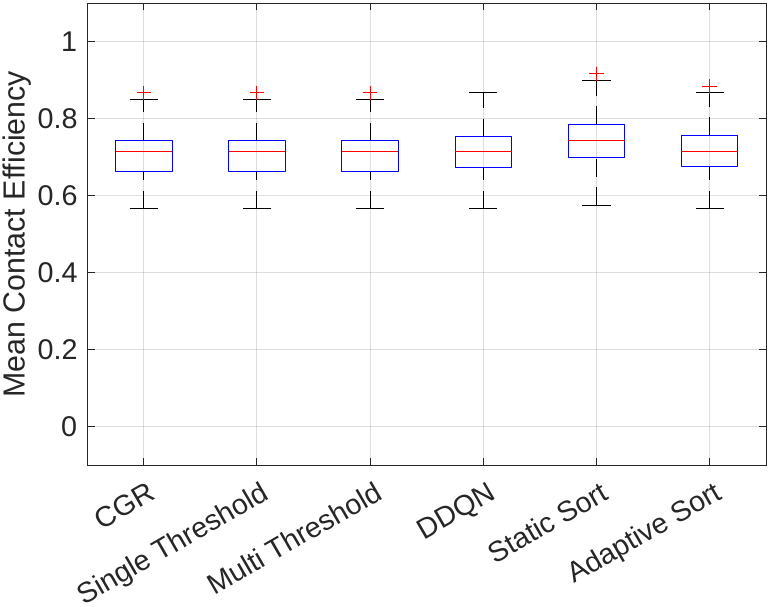}
        \subcaption{Energy efficiency when data volume is 50\% of capacity and mean cloud cover is 30\%.}
        %\subcaption{Mean contact efficiency-standard deviation 30 packets}
        \label{fig:EE_Medium_Data_Volume_Low_Cloud_Cover}
    \end{minipage}
    \caption{Results when data volume is moderate under different cloud cover conditions.}
    \label{fig:total_Medium_Data_Volume}
\end{figure*}

\subsection{Variations in Cloud Cover and Data Volume}
%Here we discuss the performance of multiple thresholds, multiple DRL agents and single of both discuss 
As discussed in Section \ref{Simulation_Config}, determining the performance of the downlink scheduling schemes under variable weather and data volume conditions is a highly relevant exercise. As per Table \ref{tab:config_table}, the first two simulation configurations shown in Figure \ref{fig:Packet_Tests} had high cloud cover conditions and low to medium data volume values. As shown in Figure \ref{fig:low_packet_del}, the static schemes generally perform poorly in this configuration, with multi-threshold completely failing to maintain the delivery ratio, and Algorithm \ref{pseudocode:Sort_Algo} showing approximately a 7\% reduction in the mean delivery ratio compared to CGR. The adaptive schemes perform significantly better in this case, with DDQN suffering only a 4\% decrease in mean delivery ratio in comparison to CGR, and adaptive sorting suffers no reduction in mean delivery ratio. 

When the data volume values are increased under the same cloud cover conditions, the energy efficiency schemes have fewer opportunities to choose more advantageous contacts, reducing mean contact energy efficiency improvements over CGR. Although the sorting algorithms successfully maintain the same mean delivery ratio as CGR, multi-threshold and DDQN suffer a 13\% reduction in performance relative to CGR. For the last two configurations specified in Table \ref{tab:config_table}, the cloud cover values were reduced to assess the performance of the downlink scheduling schemes under largely favourable conditions. In this case, the results displayed in Figure \ref{fig:total_Medium_Data_Volume} showed mixed results, with Algorithm \ref{pseudocode:Sort_Algo} and Algorithm \ref{pseudocode:Adaptive_Sort_Algo} performing slightly worse than most of the other schemes in terms of delivery ratio, while only producing small related benefits in energy efficiency. In these specific experiments, multi-threshold and DDQN perform well, maintaining delivery ratio while providing minor improvements in energy efficiency.
\subsection{\pgm{Case study}}
%Increased realism experiments
\pgm{
%Simulation introduction
Although the simulation results above provide important insights into the behaviour of the proposed downlink scheduling schemes in a controlled environment, they do not account for challenges in realistic LEO satellite systems, such as variable contact lengths, imperfect weather-forecast data, and continuous cloud-cover values. These additional factors were incorporated into the case study described in Section \ref{sec:Case_Study_Config}. The performance of the downlink scheduling schemes was again evaluated in terms of delivery ratio and mean contact energy efficiency. The resulting median, 75th percentile quartile (UQ) and 25th (LQ) percentile values are noted in Table \ref{tab:case_study_results}. Three different packet value configurations were tested, which broadly align with the simulations conducted in the previous two subsections (uniform, 0.1V and 0.5V).}

%Results
\pgm{
%Put further analysis here
In Table \ref{tab:case_study_results}, one can see that the DDQN scheme has lost significant performance when compared to previous simulations in terms of mean contact energy efficiency and delivery ratio. Showing poor performance even when compared to the simple threshold schemes.}

\pgm{
Simultaneously, due to the changes in cloud cover distribution between the training ground station sites and the testing ground station sites, the threshold schemes fail to conserve energy to the same extent as the sorting schemes. Both sorting schemes suffer some degradation in the delivery ratio, with static sorting being affected more heavily than adaptive sorting. This underlines the difficulty of setting thresholds and volume margin parameters when the ground station sites change during operations.
}
\renewcommand{\arraystretch}{1.5}
\begin{table*}
\centering
\caption{Case study results.}
\begin{tabular}{|c|c|c|c|c|c|c|c|c|c|c|}
\hline
\multicolumn{2}{|c|}{\multirow{2}{*}{\diagbox{\pgm{\bigtablefont Schemes}}{\pgm{\bigtablefont Packet config}}}}
& \multicolumn{3}{c|}{\pgm{\bigtablefont Uniform}}
& \multicolumn{3}{c|}{\pgm{ \bigtablefont 0.1V}}
& \multicolumn{3}{c|}{\pgm{\bigtablefont 0.5V}} \\
\cline{3-11}
\multicolumn{2}{|c|}{}
& \pgm{\bigtablefont median} & \multicolumn{1}{|c|}{\pgm{\bigtablefont UQ}} & \pgm{\bigtablefont LQ}
& \pgm{\bigtablefont median} & \multicolumn{1}{|c|}{\pgm{\bigtablefont UQ}} & \pgm{\bigtablefont LQ}
& \pgm{\bigtablefont median} & \multicolumn{1}{|c|}{\pgm{\bigtablefont UQ}} & \pgm{\bigtablefont LQ} \\ \hline

\multirow{2}{*}{\pgm{\bigtablefont CGR}}
& \pgm{DL} & \pgm{1.0000}  & \pgm{1.0000} &\pgm{0.6718}  & \pgm{1.0000} & \pgm{1.0000} & \pgm{1.0000} & \pgm{1.0000} & \pgm{1.0000}
& \pgm{0.8555} \\ \cline{2-11}
& \pgm{EE} & \pgm{0.5889} & \pgm{0.7139}  &\pgm{0.4204}  & \pgm{0.6250} & \pgm{0.8889} & \pgm{0.3787} & \pgm{0.5863} & \pgm{0.6830} & \pgm{0.4333} \\ \hline

\multirow{2}{*}{\pgm{\bigtablefont Single Threshold}}
& \pgm{DL} & \pgm{1.0000} & \pgm{1.0000} &\pgm{0.6564}  & \pgm{1.0000} & \pgm{1.0000} & \pgm{1.0000} & \pgm{1.0000} & \pgm{1.0000} & \pgm{0.8555} \\ \cline{2-11}
& \pgm{EE} & \pgm{0.66}  & \pgm{0.7938}  & \pgm{0.5238} & \pgm{0.6806} & \pgm{0.9167} & \pgm{0.4560} & \pgm{0.6572} & \pgm{0.7910} & \pgm{0.5490} \\ \hline

\multirow{2}{*}{\pgm{\bigtablefont Multi Threshold}}
& \pgm{DL} & \pgm{1.0000} & \pgm{1.0000}  & \pgm{0.6564}  & \pgm{1.0000} & \pgm{1.0000} & \pgm{1.0000} & \pgm{1.0000} & \pgm{1.0000} & \pgm{0.8555} \\ \cline{2-11}
& \pgm{EE} & \pgm{0.6698} & \pgm{0.8031} & \pgm{0.5028} & \pgm{0.7500} & \pgm{0.9167} & \pgm{0.5370}  &  \pgm{0.6538} & \pgm{0.7917} & \pgm{0.5111} \\ \hline

\multirow{2}{*}{\pgm{\bigtablefont DDQN}}
& \pgm{DL} & \pgm{0.8134}  & \pgm{1.0000}  & \pgm{0.5496}  & \pgm{1.0000}  & \pgm{1.0000}  & \pgm{1.0000}  & \pgm{0.8598} & \pgm{0.9233} & \pgm{0.7413} \\ \cline{2-11}
& \pgm{EE} & \pgm{0.5876} & \pgm{0.7404}  &\pgm{0.4330}  & \pgm{0.6535} & \pgm{0.8889} & \pgm{0.4151} & \pgm{0.5757} & \pgm{0.6991} & \pgm{0.4562} \\ \hline

\multirow{2}{*}{\pgm{\bigtablefont Static Sort}}
& \pgm{DL} & \pgm{0.8944} & \pgm{1.0000} & \pgm{0.6440}  & \pgm{1.0000} & \pgm{1.0000} & \pgm{1.0000} & \pgm{0.9902} & \pgm{1.0000} & \pgm{0.8209} \\ \cline{2-11}
& \pgm{EE} & \pgm{0.7183} & \pgm{0.8755} & \pgm{0.5176} & \pgm{1.0000} & \pgm{1.0000} & \pgm{0.8333} & \pgm{0.6997} & \pgm{0.8333} & \pgm{0.5176} \\ \hline

\multirow{2}{*}{\pgm{\bigtablefont Adaptive Sort}}
& \pgm{DL} & \pgm{0.9550} & \pgm{1.0000} & \pgm{0.6564} & \pgm{1.0000} & \pgm{1.0000} & \pgm{1.0000} & \pgm{1.0000} & \pgm{1.0000} & \pgm{0.8524} \\ \cline{2-11}
& \pgm{EE} & \pgm{0.7692} & \pgm{0.9013} & \pgm{0.5212} & \pgm{1.0000} & \pgm{1.0000} & \pgm{0.8333} & \pgm{0.7520} & \pgm{0.8889} & \pgm{0.5111} \\ \hline
\end{tabular}
\label{tab:case_study_results}
\end{table*}

\subsection{Discussion}
\label{discussion}
Determining the most suitable downlink scheduling scheme for a particular satellite system is influenced by several factors. Firstly, as spare computational power on satellites is limited, conducting onboard computations as required for the adaptive schemes may not be feasible, depending on the specific satellite capabilities. This results in schemes, such as the threshold and multi-threshold, as well as the Algorithm \ref{pseudocode:Sort_Algo}, being more attractive options. However, selecting these less complex schemes must be balanced with the \pgm {energy-saving} objectives and QoS requirements of the system.

Secondly, another aspect of this problem is the QoS requirements of the system. As was shown in Figure \ref{fig:low_packet_del}, both the threshold schemes and the static sorting algorithm have reduced performance in certain circumstances, particularly in terms of delivery ratio. For many systems, delivering data from the current downlink is critical, as data may be overwritten due to the satellite's limited onboard memory. In this case, adaptive techniques would provide notable benefits due to their ability to consistently deliver packets (while maximizing energy efficiency) under a wider range of data volume and weather conditions.

Thirdly, the potential variability of the ground station configuration for the satellite system is also a relevant consideration. Suppose the ground station configuration is likely to change frequently during the mission. In that case, the threshold schemes will suffer performance degradation, as tuning for each new ground station configuration might not be practical.%\pgm{This was shown in particular in the case study simulations, where threshold schemes tuned on a certain ground station configuration showed much worse performance than adaptive schemes on the test ground station configuration in terms of energy efficiency.}

%Challenges regarding transition function dynamics 
An additional challenge for the methods presented in this paper is the assumption of known transition dynamics for different cloud cover conditions. In particular, the static and adaptive sorting algorithms rely on having access to an accurate model for the expected number of packets delivered under a given set of conditions. This issue also affects the DDQN and Q-learning schemes under the current training setup, which is assumed to be conducted on the ground using a simulated environment due to the long training times. Hence, prior to deployment, the wireless channel would need to be characterized, potentially even on a site-specific basis. This challenge could be avoided using either offline reinforcement learning techniques or conducting training on the satellite in an online manner. The feasibility of the current assumptions and the alternatives depends on the nature of the ground station infrastructure and satellite capabilities.

%Case study discussion

\section{Conclusion}
\label{conclusion}
\pgm{In this paper, the downlink scheduling problem for optical delay-tolerant satellite networks was studied. This differs from previous works that primarily focused on delay-sensitive satellite networks and site-diversity-based solutions that require extensive ground infrastructure. To support this study, a system consisting of a single satellite, with link availability for each transmission opportunity modelled as a stochastic process, was defined. We showed that with a fixed downlink period, this problem can be formulated as a knapsack problem variant. To tackle this problem, downlink scheduling schemes were developed that prioritize delivery ratio while improving energy efficiency. Building on existing work in this field, adaptive strategies that account for the highly stochastic nature of the downlink scheduling problem within a downlink period were proposed.} As demonstrated in Section \ref{sec:evaluation}, adaptive schemes achieve improved performance over static schemes in several key scenarios when cloud cover distributions change. \pgm{However, it was shown that certain adaptive schemes, such as DDQN, struggle when realistic factors such as cloud cover prediction uncertainty are incorporated into the system model}. In terms of computational complexity, static schemes were found to be much less computationally intensive. Hence, the balance between computational resource requirements and overall performance is critical when determining what scheme to use on a particular satellite system.

\section{Future Work}
\label{futurework}

Several avenues of future work related to the results presented in this paper exist. As discussed in Section \ref{discussion}, the underlying assumptions of the proposed algorithm assume knowledge of the channel dynamics to a level of detail that may not be feasible in all potential circumstances. If knowledge of the channel is not available beforehand, new downlink scheduling schemes would need to be developed. With possible approaches including offline frameworks such as those presented in \cite{offline_dqn}.

Additionally, evaluating more complex satellite network architectures is another potential direction for future work. Possible extensions include expanding the problem to allow for multiple satellites and a limited number of laser terminals available at each ground station. This would be applicable to denser sparse constellations with fewer constraints on the network topology than the schemes presented in this work.
%\begin{figure}%[H]
%\centering
%\includegraphics[width=0.98\linewidth]{fig/Comparison_Baseline_Schemes/Downlink_Period_vs_Exec_time.pdf} 
%  \caption{Average execution time of the adaptive sorting algorithm with different contact lengths.}
%  \label{fig:Adapt_sort_exec_time}
%\end{figure}
%\begin{figure}%[H]
%\centering
%\includegraphics[width=0.98\linewidth]{fig/Comparison_Baseline_Schemes/average_execution_time_bar_log.pdf} 
%  \caption{Average execution time of different schemes.}
%  \label{fig:Exec_time}
%\end{figure}
\section*{Acknowledgements}
This work has been supported by the National Research Council Canada's (NRC) High Throughput Secure Networks program within the Optical Satellite Communications Consortium Canada (OSC) framework, MDA and Mitacs. \pgm{In addition, we would like to thank meteoblue for allowing access to the weather data used in this work. Grammarly was used for spelling and punctuation correction during the writing of this paper.}
%\bibliographystyle{IEEEtran}
%\bibliography{biblio.bib}

\end{document}